\documentclass[final,10pt,aps,longbibliography,twocolumn,superscriptaddress]{revtex4-1}
\usepackage{amsmath,amssymb,bm}
\usepackage{graphicx,color}
\usepackage[squaren]{SIunits}
\usepackage[english]{babel}
\usepackage{amstext}
\usepackage{amsthm}
\usepackage{latexsym}
\usepackage{array}
\usepackage{color}
\usepackage{float}
\usepackage{microtype}
 
\usepackage{multirow}
\usepackage{dcolumn}
\definecolor{url}{RGB}{0,20,160}
\usepackage[colorlinks=true,linkcolor=blue,citecolor=blue,urlcolor=url]{hyperref}
\usepackage[usenames,dvipsnames,svgnaes,table]{xcolor}

\usepackage{natbib}
\makeatletter
\def\frutiger{cmss10 }
\def\frutigerbold{cmssbx10 }
=\frutigerbold at 14pt
=\frutiger at 8pt
=\frutigerbold at 12pt
=\frutigerbold at10pt
=\frutigerbold at 8pt
=\frutiger at 8pt
=\frutigerbold at 31pt
\def\@caption@tabnum@sep{\figtextfont{{ }{\bf\textbar}{ }}}%
\def\fnum@table{{\bf\tablename~\thetable}}

\def\@caption@fignum@sep{\figtextfont{{ }{\bf\textbar}{ }}}%

\renewcommand{\fnum@figure}{\bf Fig. \thefigure}
\def\@startsection#1#2#3#4#5#6{%
\if@noskipsec\leavevmode\fi
\par\@tempskipa #4\relax
\@afterindenttrue
\ifdim\@tempskipa <\z@
\@tempskipa -\@tempskipa \@afterindentfalse
\fi\if@nobreak\everypar{}%
\else\addpenalty\@secpenalty\addvspace\@tempskipa\fi
\@ifstar{\@ssect{#3}{#4}{#5}{#6}}{\@dblarg{\@sect{#1}{#2}{#3}{#4}{#5}{#6}}}}
\def\@sect#1#2#3#4#5#6[#7]#8{%
\ifnum #2>0
\let\@svsec\@empty
\else\refstepcounter{#1}\protected@edef\@svsec{\@seccntformat{#1}\relax}\fi
\@tempskipa #5\relax
\ifdim\@tempskipa>\z@
\begingroup#6{\@hangfrom{\hskip #3\relax\@svsec}%
\interlinepenalty \@M #8\@@par}\endgroup
\csname #1mark\endcsname{#7}%
\addcontentsline{toc}{#1}{%
\ifnum #2>\c@secnumdepth\else
\protect\numberline{\csname the#1\endcsname}\fi #7}%
\else\def\@svsechd{#6{\hskip #3\relax
\@svsec #8\ifnum#2=2.\fi}%
\csname #1mark\endcsname{#7}%
\addcontentsline{toc}{#1}{%
\ifnum #2>\c@secnumdepth \else
\protect\numberline{\csname the#1\endcsname}\fi #7}}%
\fi\@xsect{#5}}
\renewcommand\section{\@startsection {section}{1}{\z@}%
{-10pt \@plus -1ex \@minus -.2ex}{.5ex }{\normalfont\Large\bfseries\sectionfont}}
\renewcommand\subsection{\@startsection{subsection}{2}{\z@}%
{10pt\@plus 1ex \@minus.2ex}{-0.5ex \@plus.2ex}{\normalfont\large\bfseries\subsectionfont}}
\def\frontmatter@title@format{\titlefont\centering}%
\def\frontmatter@title@below{\addvspace{-5pt}}%

\newcommand*\bib@heading{%
  \section{\refname}
  \fontsize{8}{10}\selectfont
}
\newcommand*\@openbib@code{%
      \advance\leftmargin\bibindent
      \itemindent -\bibindent
      \listparindent \itemindent
      \parsep \z@
}%
\newdimen\bibindent
\makeatother

\newcommand{\lbnl}{Energy Technologies Area, Lawrence Berkeley National Laboratory, Berkeley, CA 94720, USA}
\newcommand{\northwestern}{Department of Materials Science \& Engineering, Northwestern University, Evanston, IL 60208, USA}

\begin{document}

	\title{When Band Convergence is Not Beneficial for Thermoelectrics}
	\author{Junsoo Park}
	\email{qkwnstn@gmail.com}
	\affiliation{\lbnl}
	\author{Maxwell Dylla}
	\affiliation{\northwestern}
	\author{Yi Xia}
	\affiliation{\northwestern}
	\author{Max Wood}
	\affiliation{\northwestern}
	\author{G. Jeffrey Snyder}
	\email{jeff.snyder@northwestern.edu}
	\affiliation{\northwestern}
	\author{Anubhav Jain}
	\email{ajain@lbl.gov}
	\affiliation{\lbnl}
	\date{\today} 
	\begin{abstract}
Band convergence is considered a clear benefit to thermoelectric performance because it increases the charge carrier concentration for a given Fermi level, which typically enhances charge conductivity while preserving the Seebeck coefficient. However, this advantage hinges on the assumption that interband scattering of carriers is weak or insignificant. With first-principles treatment of electron-phonon scattering in CaMg$_{2}$Sb$_{2}$-CaZn$_{2}$Sb$_{2}$ Zintl system and full Heusler Sr$_{2}$SbAu, we demonstrate that the benefit of band convergence can be intrinsically negated by interband scattering depending on the manner in which bands converge. In the Zintl alloy, band convergence does not improve weighted mobility or the density-of-states effective mass. We trace the underlying reason to the fact that the bands converge at one \textbf{k}-point, which induces strong interband scattering of both the deformation-potential and the polar-optical kinds. The case contrasts with band convergence at distant \textbf{k}-points (as in the full Heusler), which better preserves the single-band scattering behavior thereby successfully leading to improved performance. Therefore, we suggest that band convergence as thermoelectric design principle is best suited to cases in which it occurs at distant \textbf{k}-points.
	\end{abstract}
	\maketitle



\section{Introduction}

Thermoelectrics represent a clean energy technology for power generation from waste heat and refrigerant-free cooling. A thermoelectric material's performance is described by its figure of merit, $zT= \frac{\alpha^{2}\sigma}{\kappa_{\text{lat}}+\kappa_{\text{e}}}T$, where $\alpha^{2}\sigma$ is the thermoelectric power factor (PF), composed of the Seebeck coefficient ($\alpha$) and electrical conductivity ($\sigma$). The total thermal conductivity is the sum of electronic thermal conductivity and lattice thermal conductivity ($\kappa_{\text{lat}}+\kappa_{\text{e}}$). A more direct metric for assessing a material's potential thermoelectric performance with optimized doping is the quality factor \cite{wangsnyderbook,materialdescriptors,qualityfactor,pbsequalityfactor}, given by $\beta\propto\frac{\mu_{w}}{\kappa_{\text{lat}}}$. Here, $\mu_{w}=\mu m_{\text{D}}^\frac{3}{2}$ is known as weighted mobility, composed of mobility ($\mu$) and the density-of-states (DOS) effective mass ($m_{\text{D}}$). The goal of designing high-$zT$ thermoelectric materials rests on two main objectives: enhancing the electronic performance, represented by $\mu_{w}$ or the PF, and reducing lattice heat dissipation represented by $\kappa_{\text{lat}}$.

Band convergence is one of the most successful approaches for systematic improvement of the electronic performance \cite{bandconvergencereview}. In essence, when multiple bands align in energy, they generate a higher carrier concentration ($n$) for a given Fermi level ($E_{\text{F}}$). This usually translates to increased $\sigma$ for given $\alpha$ enhancing the PF, or increased $m_{\text{D}}$ for given $\mu$ enhancing $\mu_{w}$. Beginning with PbTe \cite{pbtebandconvergence}, band convergence has been used to interpret the improved performance in many well-known thermoelectric compounds and alloys \cite{napbtebandconvergence,pbsebandconvergence,bisbte3bandconvergence1,bisbte3bandconvergence2,bi2te3se,cosb3bandconvergence1,cosb3bandconvergence2,getebandconvergence,mg2sibandconvergence1,highlyeffectivemg2sisn,mg2sibandconvergence2,chalcopyritebandconvergence2}. In a similar spirit, materials with inherent band multiplicity and degeneracy have been sought after for high intrinsic thermoelectric performance \cite{halfheuslerbanddegeneracy,ba2biau,heuslers,mg3sb2discovery,chalcopyritebandconvergence1,zintlorbitalengineering,euzncdsb,ybcdznsb,tafesbnatcomm,nbfesbnatcomm}. As such, a variety of descriptors and metrics for good thermoelectric materials universally promote high multiplicity of Fermi surface pockets ($N_{v}$)\cite{nolassharpgoldsmid,wangsnyderbook,materialdescriptors,complexity}. For example, weighted mobility is also often written as $\mu_{w}=\mu N_{v}^{\frac{2}{3}}m_{d}$ where $m_{d}$ is the single-pocket DOS effective mass.

\begin{figure*}[tp]\centering
\includegraphics[width=1 \linewidth]{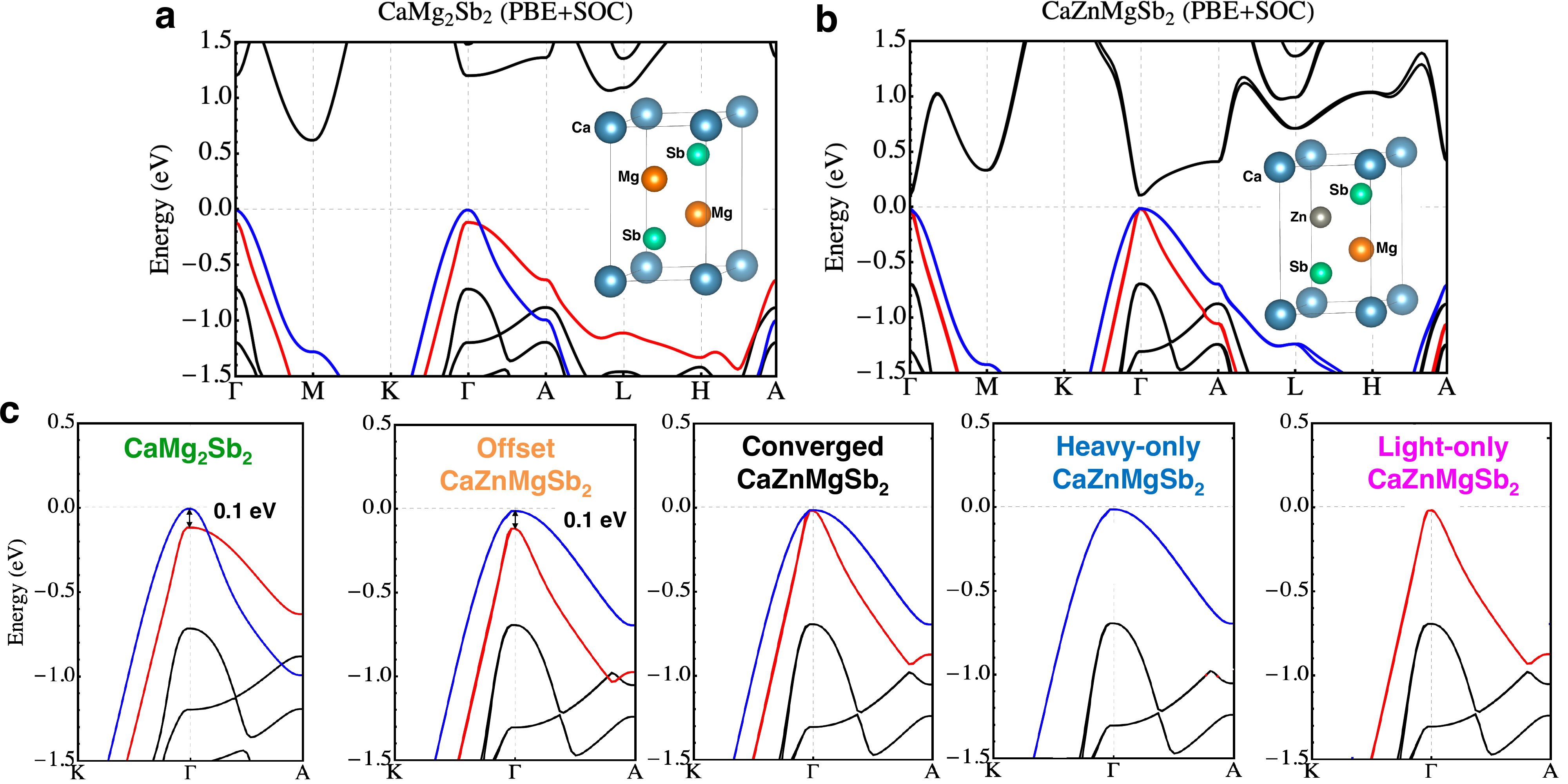}
\caption{ \textbf{DFT band structures calculated with PBE including SOC.}  \textbf{a)} CaMg$_{2}$Sb$_{2}$ and \textbf{b)} CaZnMgSb$_{2}$ band structures, with their crystal structure insets. Perfect ordering of Zn and Mg is assumed of the latter. Note that the two bands are essentially converged at $\Gamma$ for Zn:Mg = 1:1. \textbf{c)} The five band structure configurations (including hypothetical CaZnMgSb$_{2}$ configurations generated by manual adjustment of eigenvalues) subject to investigation via scattering and thermoelectric property computations.}
\label{fig:bandstructure}
\end{figure*}

Yet, such direct correlation of $N_{v}$ with improved thermoelectric performance neglects the implicit dependence of $\mu$  and $m_{\text{D}}$ on $N_{v}$ through interband (or intervalley) scattering. Experimental and theoretical analyses of multi-valley PbTe and Si suggest that although there is definite increase in $\mu_{w}$ due to pocket multiplicity, it is perhaps half of that expected purely from the number of pockets \cite{wangsnyderbook}. Other demonstrations have emerged that scattering considerations complicate the benefit of $N_{v}$ \cite{valleytronics1,valleytronics2,roleofscattering,heuslers}. Modeling based on parabolic bands has recently proposed that, under interband scattering, convergence of bands is guaranteed to be beneficial only if a lighter band converges with a heavy band \cite{bandalignmentscattering}. A related concern is the relative location of band pockets in the Brillouin zone. Band pockets that are distant in reciprocal space are typically subject to weaker scattering overall \cite{wangsnyderbook,graziosidescriptor} and, in certain cases, parity restrictions may further weaken intervalley transitions between distant symmetry-degenerate pockets\cite{pbteepwntype2}. In contrast, when multiple distinct bands are at a same \textbf{k}-point, there is little reason to presume that interband scattering would not possibly negate the benefit of band convergence.

In some experimental reports, the benefit of band convergence appears limited. Notably, in CaMg$_{2}$Sb$_{2}$, the Sb $p_{z}$ state lies above the Sb $p_{x}$ state at the valence band maximum (VBM) at the $\Gamma$-point, whereas in CaZn$_{2}$Sb$_{2}$, Sb $p_{x}$ lies above Sb $p_{z}$ to form the VBM at $\Gamma$. The $p_{z}$-state is split from $p_{x}$ and $p_{y}$ by crystal field, and the latter two are split by spin-orbit coupling (SOC). However when CaZn$_{2-x}$Mg$_{x}$Sb$_{2}$ solid solutions were synthesized, $\mu_{w}$ did not peak at the expected point of full band convergence at $x=0.86$ \cite{zintlvalencecrossing}. Other than for the Mg-end ($x=2$), both the PF and $\mu_{w}$ essentially plateaued for all $x$ all the way to the the Zn-end ($x=0$). Alloy disorder scattering of carriers was suggested to be one culprit, but it alone cannot account for it since band convergence is often achieved through alloying. This warrants investigations at a more fundamental level of the scattering behaviors and thermoelectric response with respect to band convergence. 

We herein investigate if the lack of performance increase could be attributable to an intrinsic process, namely electron-phonon (e-ph) scattering, including deformation-potential scattering (DPS) and polar-optical scattering (POS). We perform first-principles e-ph scattering and transport computations on CaZn$_{2-x}$Mg$_{x}$Sb$_{2}$ Zintl alloys as well as several hypothetical modifications of this band structure. Our results indicate that interband e-ph scattering and its changing behavior with band convergence are indeed inherently responsible and sufficient to explain the stagnant PF and $\mu_{w}$ in the Zintl alloys. We also highlight the critical role played by the relative locations of bands within a Brillouin zone: band convergence or multiplicity at distant $\textbf{k}$-points is more promising than it occurring at the same $\textbf{k}$-point, by design. In emphasis, this study strictly concerns the electronic performance represented by the PF and $\mu_{w}$, not $zT$ which may also benefit from reduced $\kappa_{\text{lat}}$ via alloying for band convergence. Also, we interpret ``band convergence" strictly as shifts in band energies, \textit{i.e.}, closing of energy offset between two bands of interest, independent of how they may also change in shapes during alloying in experiments. Note that where convenient and unambiguous, we use the Hartree atomic units ($\hbar=e=m_{e}=4\pi\epsilon_{0}=1$).

\begin{figure*}[tp]\centering
\includegraphics[width=1\linewidth]{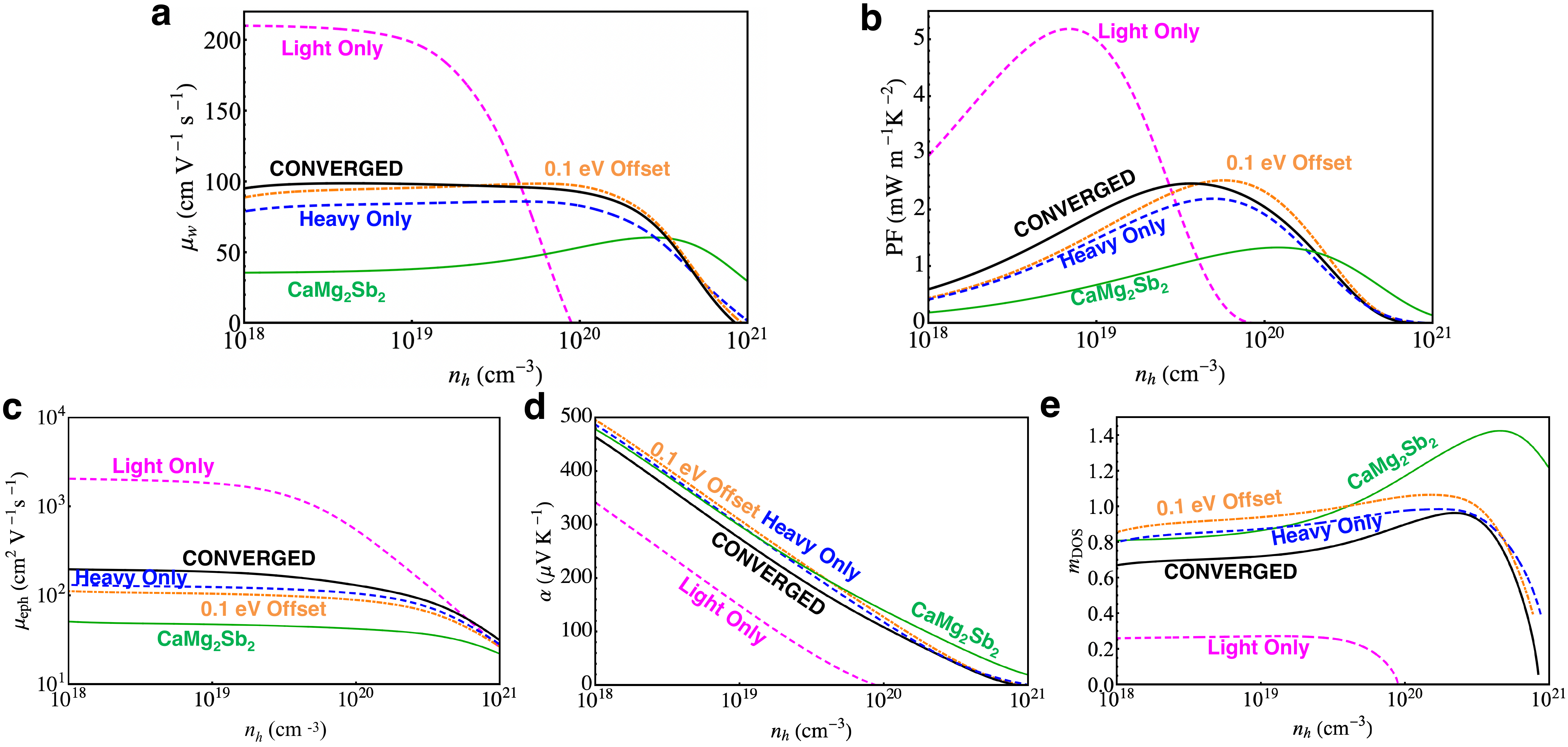}
\caption{ \textbf{Calculated e-ph limited $p$-type properties of the band configurations depicted in Fig. \ref{fig:bandstructure}c at 600 K.} \textbf{a)} Weighted mobility, \textbf{b)} the power factor, \textbf{c)} mobility, \textbf{d)} the Seebeck coefficient, and \textbf{e)} the DOS effective mass.}
\label{fig:electronic}
\end{figure*}

\section{Results}

We begin by theoretically confirming the occurrence of band convergence in CaZn$_{2-x}$Mg$_{x}$Sb$_{2}$ solid solutions. We calculate density-functional theory (DFT) \cite{dft} electronic structures including SOC using Vienna \textit{Ab initio} Simulation Package (VASP) \cite{vasp1,vasp2,vasp3,vasp4} with projector-augmented wave (PAW) pseudopotential \cite{paw} and Perdew-Burke-Ernzerhof (PBE) exchange-correlation functional \cite{pbe}. We perform the calculations on the midway alloy, CaZnMgSb$_{2}$, as well as CaMg$_{2}$Sb$_{2}$, an end compound (as explained in Supplementary Fig. S2, the CaZn$_{2}$Sb$_{2}$ endpoint is omitted). As seen in Fig. \ref{fig:bandstructure}a--b, the two valence bands essentially converge in energy for $1:1$ ratio of Zn and Mg, closing their 0.096 eV offset (hereafter rounded to 0.1 eV) in CaMg$_{2}$Sb$_{2}$. These results suggest that the bands likely did converge in the experiments for the alloy (albeit at $x=0.86$), but still failed to translate to higher thermoelectric performance. Of note, the upper band is heavier than the lower band in CaZnMgSb$_{2}$, and it is the light lower band that rises to converge on the heavy upper band as $x$ decreases. The upper band has band effective mass (inverse curvature) $m\approx0.4$, and the lower band has $m\approx0.1$.

To probe the effect of band convergence, we compare the performance of the converged alloy with respect to the CaMg$_{2}$Sb$_{2}$ endpoint as well as several hypothetical band structures. As summarized in Fig. \ref{fig:bandstructure}c, they include the CaZnMgSb$_{2}$ with a band offset enforced to be the same value as that of CaMg$_{2}$Sb$_{2}$ (but retaining the band shapes of CaZnMgSb$_{2}$); CaZnMgSb$_{2}$ with no band offset, \textit{i.e.}, fully converged bands (nearly the case from DFT with no adjustment); CaZnMgSb$_{2}$ with the light band practically removed (offset by 1 eV and rendered irrelevant); and CaZnMgSb$_{2}$ with the heavy band practically removed (offset by 1 eV and rendered irrelevant). 

\begin{figure*}[tp]\centering
\includegraphics[width=1 \linewidth]{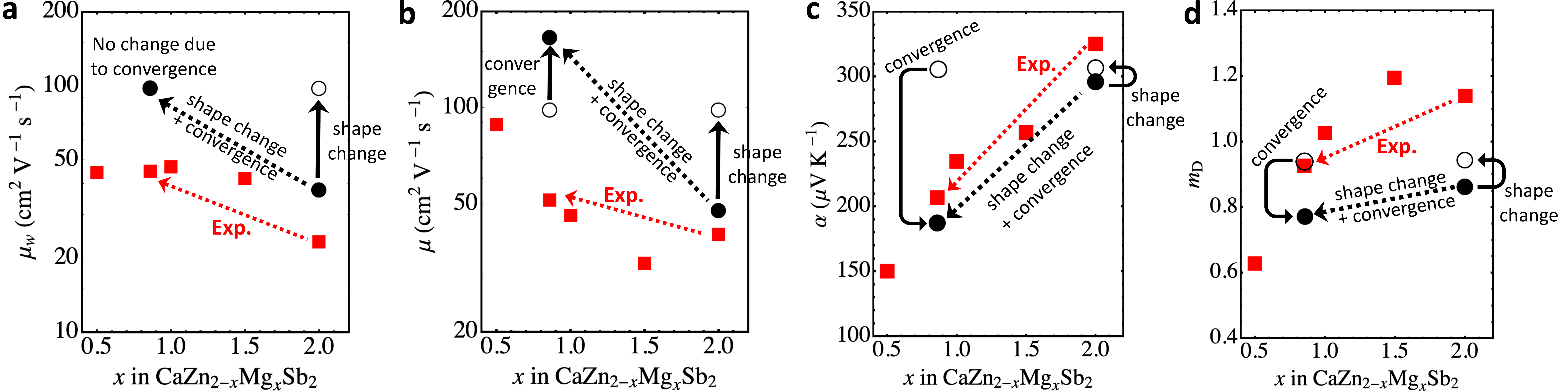}
\caption{ \textbf{Computational explanation of the experimental trends with alloy composition.} \textbf{a)} Weighted mobility, \textbf{b)} mobility, \textbf{c)} the Seebeck coefficient and \textbf{d)} the DOS effective mass. The temperature is 600 K. The experimental data points from Ref. \onlinecite{zintlvalencecrossing} are \color{red}red \color{black} squares. The calculated values are black circles. The two solid black circles correspond to CaMg$_{2}$Sb$_{2}$ at $x=2$, and the band-converged CaZnMgSb$_{2}$, which we place at $x=0.86$ (the experimental composition of band convergence). The open circle is CaZnMgSb$_{2}$ with 0.1 eV offset between the bands, which we place at both $x=2$ and $x=0.86$. The computational values track the experimental trends in two steps. First, the change from CaMg$_{2}$Sb$_{2}$ to the offset CaZnMgSb$_{2}$ is purely attributable to changes in band shapes. Then, the change from the offset CaZnMgSb$_{2}$ to the band-converged CaZnMgSb$_{2}$ is purely attributable to band convergence. Quantitative discrepancies between calculation and experiment are largely attributable to grain boundary scattering and disorder scattering; see supplementary Fig. S4.}
\label{fig:electronicvsx}
\end{figure*}

Figs. \ref{fig:electronic}a--b demonstrate that, whether measured by $\mu_{w}$ or the peak PF, the band-converged configuration is not the best design target, even with a lighter band converging onto a heavier band. The fully converged scenario is far outperformed by the hypothetical light-band-only case, matches the performance with 0.1 eV offset, and manages to slightly outperform the hypothetical heavy-band-only case. The light-band-only case is most desirable owing to its highest $\mu$, as seen in Fig. \ref{fig:electronic}c. The most interesting comparison is between the full band-converged case and the two bands offset by 0.1 eV. The former yields higher $\mu$ due to the convergence of the lighter band, but lower $\alpha$ and $m_{\text{D}}$, as shown in Figs. \ref{fig:electronic}d--e. As a result, neither the PF nor $\mu_{w}$ benefits from closing the $0.1$ eV offset. These behaviors visually summarized in Fig. \ref{fig:electronicvsx} with experimental comparisons.

\begin{figure*}[tp]\centering
\includegraphics[width=1 \linewidth]{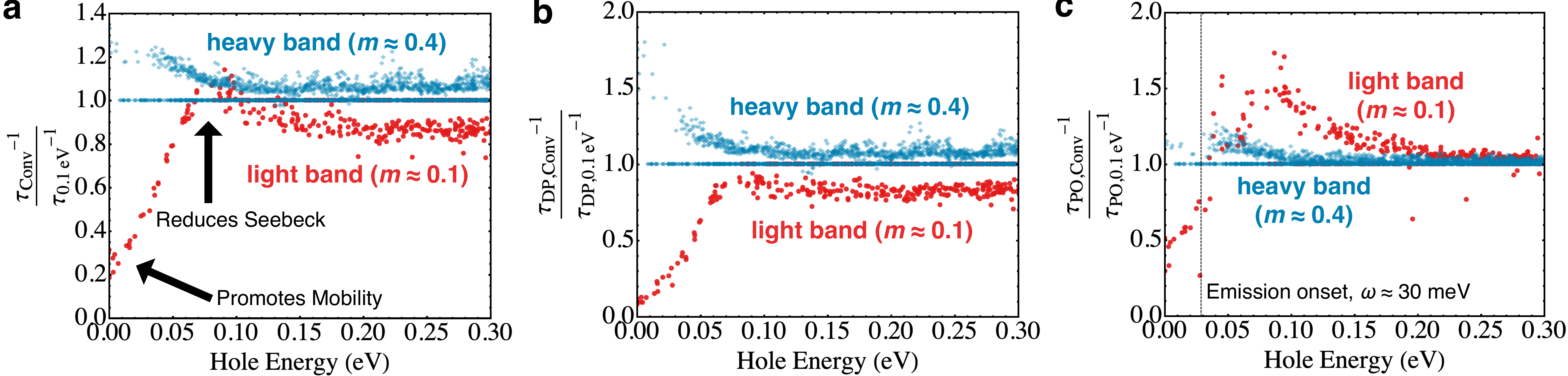}
\caption{ \textbf{Scattering rate ratios (converged versus 0.1 eV offset) of CaZnMgSb$_{2}$ at 600 K}. A value above (below) unity indicates that scattering increases (decreases) due to convergence. The energy scale is zero-referenced to the VBM. \textbf{a)} The overall scattering rate ratio. \textbf{b)} The DPS ratio, whereby the upper band only increases in scattering, while the lower band only decreases in scattering. \textbf{c)} The POS ratio, whereby both bands increase in scattering past the emission onset.}
\label{fig:eph}
\end{figure*}

The responsible physics for the behavior described above lies with the altered scattering behaviors due to band convergence, as demonstrated in Fig. \ref{fig:eph}. First, interband scattering inherently limits $\mu$-enhancement. As the lower lighter band converges with the heavier upper band, overall scattering generally decreases for the lighter band but increases for the upper band (see Fig. \ref{fig:eph}a). The reason is that the lighter, lower band loses phase space as it is pushed towards the band edge whereas the upper band gains phase space for scattering as the lower band is introduced. The relative changes in the phase spaces in turn stems from the changing local total DOS as the bands converge. We indeed find that the scattering rates largely scale as the total DOS as expected from DPS (see Fig. S3 in the SI). The overall $\mu$ still increases somewhat because the light lower band, which has higher group velocity, grows in population and lifetimes during convergence, contributing more to charge conduction. Nevertheless in the presence of interband scattering, $\mu$ does not improve as much as it could in the absence of interband scattering. We note that in real samples, external mechanisms such as grain-boundary and disorder scattering further damage mobility \cite{chalcogenidegrainboundary,mg2sigrainboundary,mg3sb2grainboundary,snsegrainboundary,srtio3grainboundary,hgtedisorderscattering}. These discussions, as well as calculated temperature-dependent transport properties compared with experiment, are provided in Figs. S4-5 and Supplementary Discussion.

Second, the reduction of $\alpha$ can be traced to the portion of the lower band that increases in scattering due to convergence, at about 0.08 eV into the band in Fig. \ref{fig:eph}a. Increased scattering for the lower band due to convergence is not an allowed behavior under DPS as verified by Fig. \ref{fig:eph}b. We attribute this to POS as verified by Fig. \ref{fig:eph}c. POS is capable of increasing for both bands as they converge, past the emission onset. Because only the optical phonons near $\Gamma$ ($\mathbf{q}\approx0$) generate strong electric fields, POS characteristically intensifies with the proximity of the initial and the final states in the reciprocal space, as described by $\tau^{-1}_{\text{POS}}\propto \left|\mathbf{k}-\mathbf{k^{\prime}}\right|^{-2}$ \cite{ziman,lundstrom}. Because energy surfaces of the two bands close on each other during convergence, POS increases for both bands, and about 0.08 eV into the light band, it overpowers DPS for increased overall scattering. For the light band states near the band edge that are below the emission onset, POS decreases just like DPS. Increasing (decreasing) scattering at high (low) energies reduces $\alpha$. Reduction of $\alpha$ in turn results in reduction of $m_{\text{D}}$ via Eq. \ref{eq:dosmass}. In all, these effects negate the minor improvement in $\mu$ leading to stagnation of $\mu_{w}$ and the PF. 

We stress that the scattering behaviors observed in Fig. \ref{fig:eph} are qualitatively reproducible using realistic model band structures and scattering models of Ref. \citenum{optimalbandstructure}. Reduction of $\alpha$ and $m_{\text{D}}$ due to band convergence with changing interband scattering is also reproducible by the same account. These more detailed explanations for the behavior of $\alpha$ and $m_{\text{D}}$ are given in the SI (see Figs. S6--S8) and establish the generality of our findings for the Zintl alloys. 

\begin{figure*}[tp]\centering
\includegraphics[width=0.8 \linewidth]{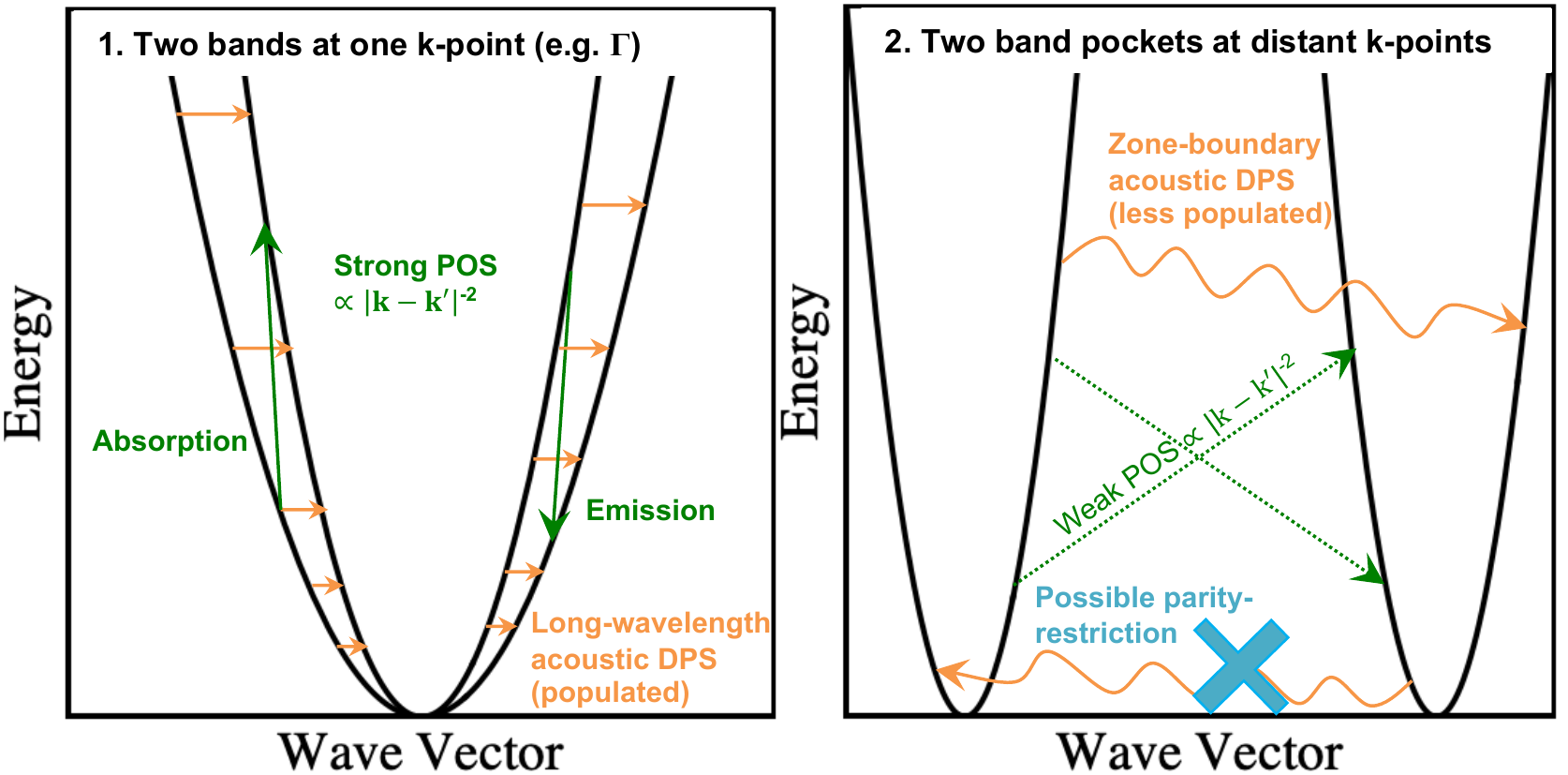}
\caption{ \textbf{Schematic of e-ph interband scattering mechanisms.} They render intervalley scattering generally stronger between bands at one \textbf{k}-point than between bands at distant \textbf{k}-points.}
\label{fig:cases}
\end{figure*}

An underlying problem for these Zintl alloys is that the bands are converging at one \textbf{k}-point, allowing 1) zone-center acoustic phonons (near $\Gamma$) to incur interband DPS, and 2) interband POS to strongly manifest. Zone-center acoustic modes are some of the most populated. Zone-boundary phonons capable of coupling states across distant \textbf{k}-points in the Brillouin zone are of higher energies and less populated. Moreover, states located at distant \textbf{k}-points virtually cannot be coupled via POS owing to the relation $\tau^{-1}_{\text{POS}}\propto\left|\mathbf{k}-\mathbf{k^{\prime}} \right|^{-2}$, whereas those located in proximity of one another are easily coupled. Distinctively, as discussed above, interband POS reduces $\alpha$ because it preferentially scatters higher energy states of both bands during convergence. Of note, ionized-impurity scattering (IIS) similarly cannot couple distant states because $\tau_{\text{IIS}}^{-1}\propto\left(\left|\mathbf{k}-\mathbf{k^{\prime}}\right|^{2}+\gamma^{2}\right)^{-2}$ where $\gamma$ is inverse screening distance \cite{ionizedimpurity,graziosidescriptor}. Furthermore, in systems with inversion symmetry, it is known that interband DPS between symmetry-degenerate band pockets could be prohibited, for any phonon mode, by parity relations \cite{pbteepwntype1}. Interband scattering between two distinct bands at one \textbf{k}-point, though possibly prohibited for certain phonon modes by orbital symmetry, is generally allowed. All things considered, band convergence at one \textbf{k}-point is not as promising by design as that at distant \textbf{k}-points. Fig. \ref{fig:cases} summarizes inherent factors that render interband scattering of various kinds between states at faraway \textbf{k}-points generally weaker than those between states at one \textbf{k}-point.

\begin{figure*}[tp]\centering
\includegraphics[width=1 \linewidth]{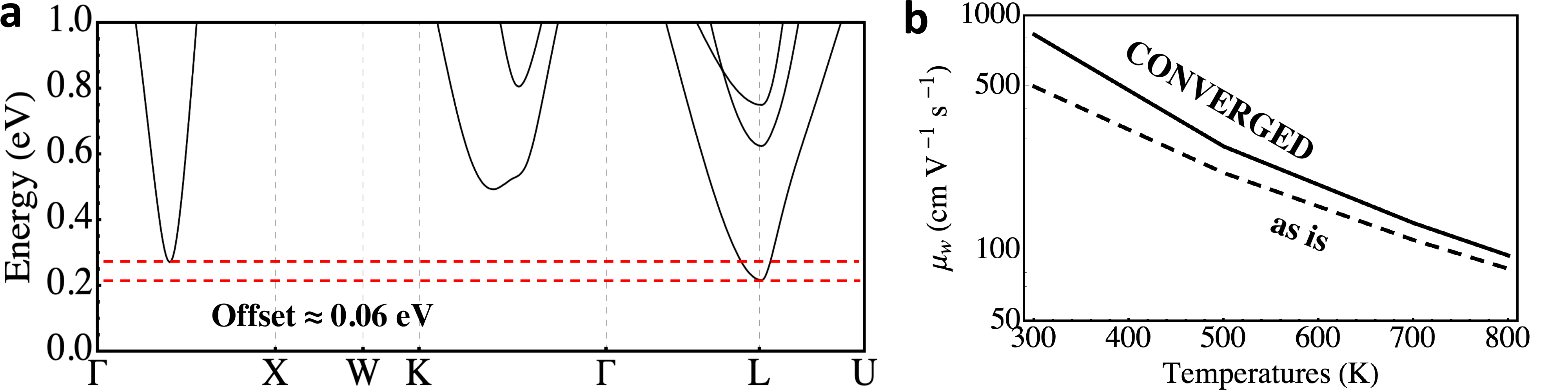}
\caption{ \textbf{Full-Heusler Sr$_{2}$SbAu, an example of band convergence at distant \textbf{k}-points.} \textbf{a)} The conduction band structure calculated with PBE including SOC. The dotted horizontal red lines indicate the minima of the $L$-pocket and the $\Gamma-X$ pocket. \textbf{b)} The weighted mobility with and without band convergence.}
\label{fig:sr2sbau}
\end{figure*}

To support the above hypothesis and draw a contrast to the Zintl scenario, we also perform a case study of Sr$_{2}$SbAu, a full-Heusler compound predicted to have high $n$-type performance \cite{heuslers} and in which (near) band convergence occurs at distant \textbf{k}-points. The conduction bands of Sr$_{2}$SbAu are depicted in Fig. \ref{fig:sr2sbau}a. The highly dispersive sixfold $\Gamma-X$-pocket has been found to be limited by POS, and is minimally affected by intervalley scattering with the fourfold $L$-pocket \cite{heuslers}. The two pockets are originally offset by approximately 0.06 eV, with the heavier $L$-pocket being lower in energy. We take the same computational approach taken for the Zintl alloy, and enforce full band convergence by pushing the $L$-pockets up by 0.06 eV. As expected from our discussion, when the bands converge, $\mu_{w}$ improves across all temperatures due to the valleys being distant in reciprocal space, as seen in Fig. \ref{fig:sr2sbau}b.

\section{Discussion}

We stress that, experimentally, it is often difficult to quantify exactly how much benefit to performance is provided by band convergence alone because the effect of band convergence is never systematically isolated in experiments. Band convergence is typically achieved via alloying, whose process is expected to alter the band curvatures as well rather than purely achieving energy convergence. Such band-shape alteration occurs in Mg$_{3}$Sb$_{2-x}$Bi$_{x}$, which is found to be responsible for $\mu_{w}$ improvement in the alloys \cite{mg3sb2bandengineering}. Likewise in this work, the band shapes change going from CaMg$_{2}$Sb$_{2}$ to CaZnMgSb$_{2}$. This, \textit{not} band convergence, is responsible for the initial performance improvement in the dilute-Zn regime coming off of the CaMg$_{2}$Sb$_{2}$-end, as can be read by comparing it to CaZnMgSb$_{2}$ with the same offset in Fig. \ref{fig:electronic}. As $x$ decreases further, however, the two bands approach in energy but without improvement in $\mu_{w}$, as evidenced by both the experimental results in Ref. \onlinecite{zintlvalencecrossing} and our computational results. Furthermore, experimental reports on benefits of band convergence were often established by means of the PF or $zT$ rather than $\mu_{w}$. The PF is a doping-dependent quantity unlike $\mu_{w}$ \cite{weightedmobility}, which may not have been optimized in each and every case, and $zT$ may simply benefit from concomitantly lowered $\kappa_{\text{lat}}$ in alloys, which is besides our point.

For clarification, our study does not mean that band convergence at one \textbf{k}-point can never be of benefit. If interband scattering is not as strong as what we find for CaZn$_{2-x}$Mg$_{x}$Sb$_{2}$, band convergence even at one \textbf{k}-point could provide intrinsic benefit. However, our results and the underlying physics suggest that band convergence and multiplicity are much more likely to offer stronger benefits when convergence occurs at distant \textbf{k}-points. With hindsight, it is perhaps not a surprising corollary that hallmark cases of successful band convergence have generally occurred at distinct \textbf{k}-points. Filled skutterudite Y$_{x}$Co$_{4}$Sb$_{12}$ conduction bands have the twelvefold pocket along $\Gamma-N$ converging with the triply degenerate band minimum at the $\Gamma$-point for improved $n$-type performance \cite{cosb3bandconvergence1,cosb3bandconvergence2}. In $p$-type Bi$_{2-x}$Sb$_{x}$Te$_{3}$ alloys, a threefold pocket along $\Gamma-Z$ and another along $Z-F$ converge \cite{bisbte3bandconvergence1,bisbte3bandconvergence2}. In $p$-type PbTe, where band convergence was first demonstrated, two valence band pockets at the fourfold $L$-point and along the twelve-fold $\Sigma$-line get close to convergence \cite{pbtebandconvergence}. Moreover, it has recently been determined based on first-principles calculations that POS is the dominant mechanism for the PbTe valence bands, which would be weak between the distant pockets \cite{pbteepwptype,pbtetdependentbandstructure}. Similar effect is observed in PbSe \cite{pbsebandconvergence}. That said, Mg$_{2}$Si$_{1-x}$Sn$_{x}$ would warrant an in-depth study, whose alloys display higher PF than either end compound and undergo band convergence at one \textbf{k}-point at some intermediate $x$ \cite{highlyeffectivemg2sisn,mg2sibandconvergence1,mg2sibandconvergence2}. In Fig. S9 and Supplementary Discussions, we confirm that the bands do converge for the alloy by calculating its band structure with BandUP \cite{bandup1,bandup2} and HSE06 hybrid functional \cite{hse1,hse2}, but discuss that it remains unclear whether it is indeed responsible for the peak performance at intermediate $x$.

In conclusion, band convergence does not always improve thermoelectric performance. We show theoretically that intrinsic electron-phonon scattering between bands is sufficient for rendering band convergence ineffective in Zintl CaZn$_{2-x}$Mg$_{x}$Sb$_{2}$ alloys. Interband transitions are mediated by the highly populated zone-center acoustic phonons as well as long-wavelength polar optical phonons. During convergence, the former limits mobility enhancement by increasing scattering for the upper band while the latter lowers the Seebeck coefficient by simultaneously increasing scattering for the converging band. We find that these phenomena are by design much more likely if bands converge at the same \textbf{k}-point. Dispersive pockets at distant \textbf{k}-points tend to be less susceptible to strong intervalley scattering; such transitions require less populated zone-boundary phonons and virtually cannot be coupled by polar optical phonons. Therefore, convergence at distant \textbf{k}-points generally appears much more promising than that at one \textbf{k}-point by design. This study should offer additional layer of guidance for experimentalists in their rational design and optimization of high-performing thermoelectrics. In particular, the weighted mobility relation $\mu_{w}=\mu N_{v}^{\frac{2}{3}}m_{d}$ may need to be reconsidered in a manner that accounts for the negative correlation between $\mu$, $m_{d}$ and $N_{v}$ via interband scattering.

\section*{Methods}

Electron-phonon scattering rates are derived from the imaginary part of electron self-energies, and ultimately calculated as
\begin{widetext}
\begin{equation}\label{eq:eph}
\begin{aligned}
\tau^{-1}_{\nu\mathbf{k}}=\frac{2\pi}{N_{\mathbf{q}}}\sum_{\nu'\lambda\mathbf{q}}
\left| g_{\nu^{\prime}\nu\lambda\mathbf{k}\mathbf{q}} \right|^{2} [ & (b(\omega_{\lambda\mathbf{q}},T)+f(E_{\nu'\mathbf{k}+\mathbf{q}},E_{\text{F}},T))\delta(E_{\nu\mathbf{k}}+\omega_{\lambda\mathbf{q}}-E_{\nu'\mathbf{k}+\mathbf{q}}) \\
& + (b(\omega_{\lambda\mathbf{q}},T)+1-f(E_{\nu'\mathbf{k}+\mathbf{q}},E_{\text{F}},T))\delta(E_{\nu\mathbf{k}}-\omega_{\lambda\mathbf{q}}-E_{\nu'\mathbf{k}+\mathbf{q}}) ],
\end{aligned}
\end{equation}
\end{widetext}
where $g_{\nu^{\prime}\nu\lambda\mathbf{k}\mathbf{q}}$ represents the e-ph interaction (scattering) matrix elements for coupling between electronic states $\nu^{\prime}$ with wavevector \textbf{k}+\textbf{q} and $\nu$ with wavevector \textbf{k} due to phonon mode $\lambda$ of wavevector \textbf{q} and frequency $\omega$ Also, $b$ and $f$ are, respectively, the Bose-Einstein and Fermi-Dirac distributions for phonons and electrons, while $\delta$ is energy-and-momentum-conserving delta function.

To compute Eq. \ref{eq:eph}, we use the EPW software \cite{epw1,epwpolar,epw3}, which interpolates coarse-mesh electronic states, phonon states, and e-ph interaction matrix elements onto dense \textbf{k}-point (for electrons) and \textbf{q}-point (for phonons) meshes using maximally localized Wannier functions \cite{mlwfcomposite,mlwfentangled,wannier90}. The coarse-mesh matrix elements are calculated using density functional perturbation theory (DFPT) \cite{ponceprb,poncejcp,epwreview} as implemented in Quantum Espresso \cite{qespresso1,qespresso2} using Optimized Norm-Conserving Vanderbilt (ONCV) pseudopotentials \cite{oncv1,oncv2,oncv3} with the PBE functional. We calculate on coarse meshes of $8\times8\times4$ and $4\times4\times2$ and interpolate onto dense meshes of $80\times80\times60$ and $40\times40\times30$ for \textbf{k} and \textbf{q}, respectively. We calculate both deformation-potential and polar-optical scattering rates, whose matrix elements are given in Supplementary Methods.

Critically, we probe the effects of band convergence by manually shifting the interpolated eigenenergies of the bands prior to phase-space integration. We do so as to 1) force exact convergence at the $\Gamma$-point, 2) maintain the 0.1 eV energy offset (inherited from CaMg$_{2}$Sb$_{2}$), 3) simulate a hypothetical case where only the heavy upper band exists, and 4) simulate another hypothetical case where only the light lower band exists (see Fig. 1c for in the main text for the schematic). In strict terms, 3) and 4) actually involves offsetting the respective bands by full 1 eV, which essentially renders them irrelevant for thermoelectric transport while still maintaining the total number of electrons. These interventions thereby modulates only the $\delta$-functions in Eq. \ref{eq:eph}, offering a pure phase-space contrast between the fully band-converged and other cases, while keeping all other components identical, namely the band shapes.

Thermoelectric properties are computed by Boltzmann transport integrals using the BoltzTraP code \cite{boltztrap}, which we have modified in such a way that it takes band-and-\textbf{k}-dependent $\tau$ as inputs, in the same format as electronic eigenenergies, and then Fourier-interpolates them just as it does electronic eigenenergies. Weighted mobility is obtained from $\sigma$ and $\alpha$ \cite{weightedmobility},
\begin{widetext}

\begin{equation}\label{eq:weightedmobility}
\mu_{w}=\frac{3\pi^{2}\sigma}{(2k_{\text{B}}T)^{\frac{3}{2}}}\left[ \frac{\text{exp}\left(\frac{|\alpha|}{k_{\text{B}}}-2\right)}{1+\text{exp}\left(-5\left(\frac{|\alpha|}{k_{\text{B}}}-1\right)\right)} + \frac{\frac{3|\alpha|}{\pi^{2}k_{\text{B}}}}{1+\text{exp}\left(5\left(\frac{|\alpha|}{k_{\text{B}}}-1\right)\right)} \right],
\end{equation}

by which account the DOS effective mass is 

\begin{equation}\label{eq:dosmass}
m_{\text{D}}=\left(\frac{3\pi^{2}n}{(2k_{\text{B}}T)^{\frac{3}{2}}}\left[ \frac{\text{exp}\left(\frac{|\alpha|}{k_{\text{B}}}-2\right)}{1+\text{exp}\left(-5\left(\frac{|\alpha|}{k_{\text{B}}}-1\right)\right)} + \frac{\frac{3|\alpha|}{\pi^{2}k_{\text{B}}}}{1+\text{exp}\left(5\left(\frac{|\alpha|}{k_{\text{B}}}-1\right)\right)} \right]\right)^{\frac{2}{3}},
\end{equation}

\end{widetext}
where $k_{\text{B}}$ is Boltzmann's constant and $n$ is carrier concentration. These relations render $\mu_{w}$ and $m_{\text{D}}$ doping-independent in the non-degenerate regime so long as scattering is also doping-independent \cite{weightedmobility}. For transport calculations, we use band gaps determined with the Tran and Blaha's modifed Becke-Johnson (mBJ) potential \cite{mbj,zintlmbjcrta}, which are 1.12 eV CaMg$_{2}$Sb$_{2}$ and 0.33 eV for CaZn$_{2}$Sb$_{2}$. For the alloy, we use the Vegard's-law-interpolated value of 0.66 eV for the composition CaZn$_{1.14}$Mg$_{0.86}$Sb$_{2}$. For reference, the mBJ gap directly calculated for CaZnMgSb$_{2}$ is 0.80 eV. Supplementary Methods has further details. 

While transport properties are calculated across a wide range of carrier concentrations ($10^{18}\sim10^{21}$) cm$^{-3}$, we calculate and use $\tau^{-1}$ with the Fermi level fixed at 0.05 eV away from the valence band maximum. In practice, e-ph scattering rates and lifetimes are usually not computed at every carrier concentration of interest because, unless doping is heavily degenerate, $\tau^{-1}$ has negligible dependence on the carrier occupation terms ($f$) in Eq. \ref{eq:eph} as phonon occupation ($b$) is usually far larger (especially so for acoustic modes).

\section*{Acknowledgements}
This work was intellectually led by the U.S. Department of Energy, Office of Basic Energy Sciences, Early Career Research Program, which funded JP and AJ Lawrence Berkeley National Laboratory is funded by the Department of Energy under award DE-AC02-05CH11231. This work used resources of the National Energy Research Scientific Computing Center, a Department of Energy Office of Science User Facility supported by the Office of Science of the U.S. Department of Energy under Contract No. DE-AC02-05CH11231. GJS, MD, and MW acknowledge NSF DMREF award \#1729487.

\section*{Author Contributions}
JP primarily designed the study and performed the computations under supervision of AJ. MD, YX, MW and GJS helped design the study, interpret results, and draft the manuscript.

\section*{Data Availability}

The data that support the findings of this study are available from the corresponding author upon reasonable request.

\section*{Code Availability}

The custom codes used to perform the study are available from the corresponding author upon reasonable request.

\section*{Conflicts of Interest}

There is no conflict of interest to declare.

\bibliography{references}

\end{document}


\title{Supplementary Information: When Band Convergence is Not Beneficial for Thermoelectrics}
	\author{Junsoo Park}
	\email{qkwnstn@gmail.com}
	\affiliation{\lbnl}
	\author{Maxwell Dylla}
	\affiliation{\northwestern}
	\author{Yi Xia}
	\affiliation{\northwestern}
	\author{Max Wood}
	\affiliation{\northwestern}
	\author{G. Jeffrey Snyder}
	\email{jeff.snyder@northwestern.edu}
	\affiliation{\northwestern}
	\author{Anubhav Jain}
	\email{ajain@lbl.gov}
	\affiliation{\lbnl} 
	\date{\today} 
	\maketitle

\section{Supplementary Figures}
	
\begin{figure}[hp]
\includegraphics[width=0.7 \linewidth]{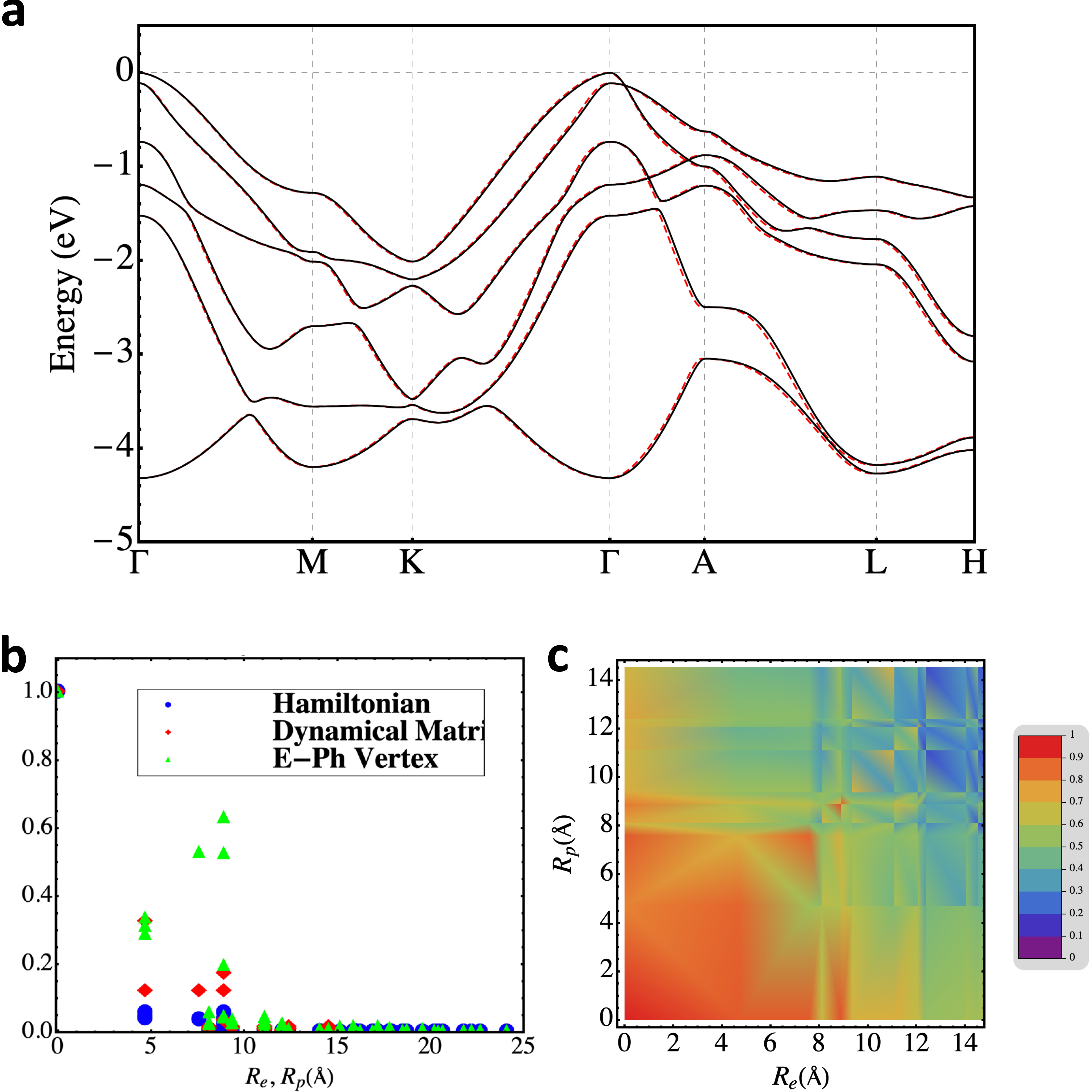}
\caption{ \textbf{Wannier interpolation quality check.} \textbf{a)} Wannier-interpolated valence band structure (dotted red) excellently reconstructs DFT-calculation (black). \textbf{b)} Decay of interpolated quantities in the Wannier representation: decay of Hamiltonian and the e-ph matrix elements are with respect to electronic distance ($R_{e}$), and decay of the dynamical matrix is with phonon distance ($R_{p}$). \textbf{c)} Decay of the e-ph matrix elements in electronic and phonon distances.}
\label{fig:decay}
\end{figure}

\newpage

\begin{figure}[hp]
\includegraphics[width=0.7 \linewidth]{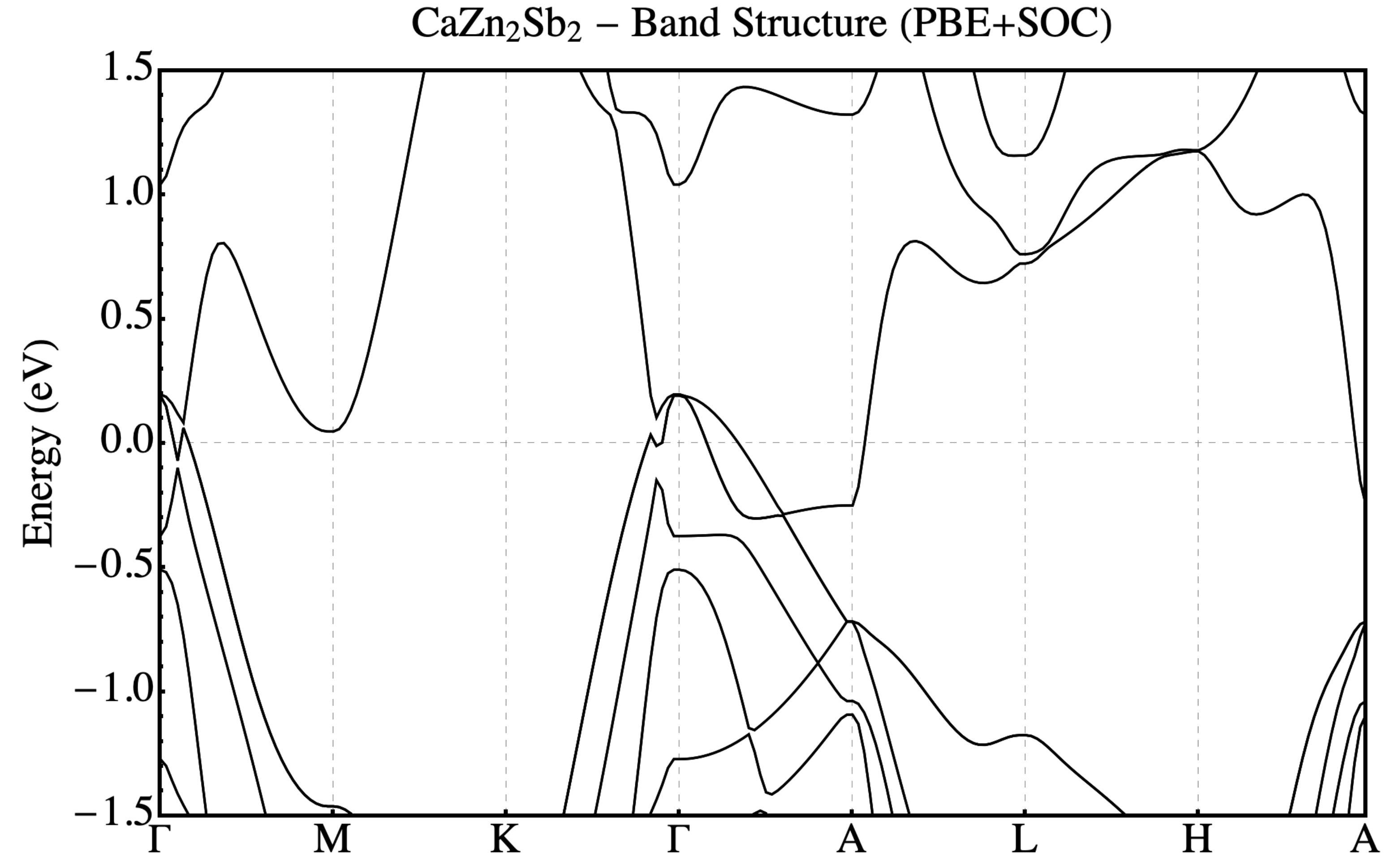}
\caption{ \textbf{The PBE+SOC band structure of CaZn$_{2}$Sb$_{2}$}. It is incorrectly metallic and not amenable to EPW study.}
\label{fig:cazn2sb2}
\end{figure}

\newpage

\begin{figure}[hp]
\includegraphics[width=1 \linewidth]{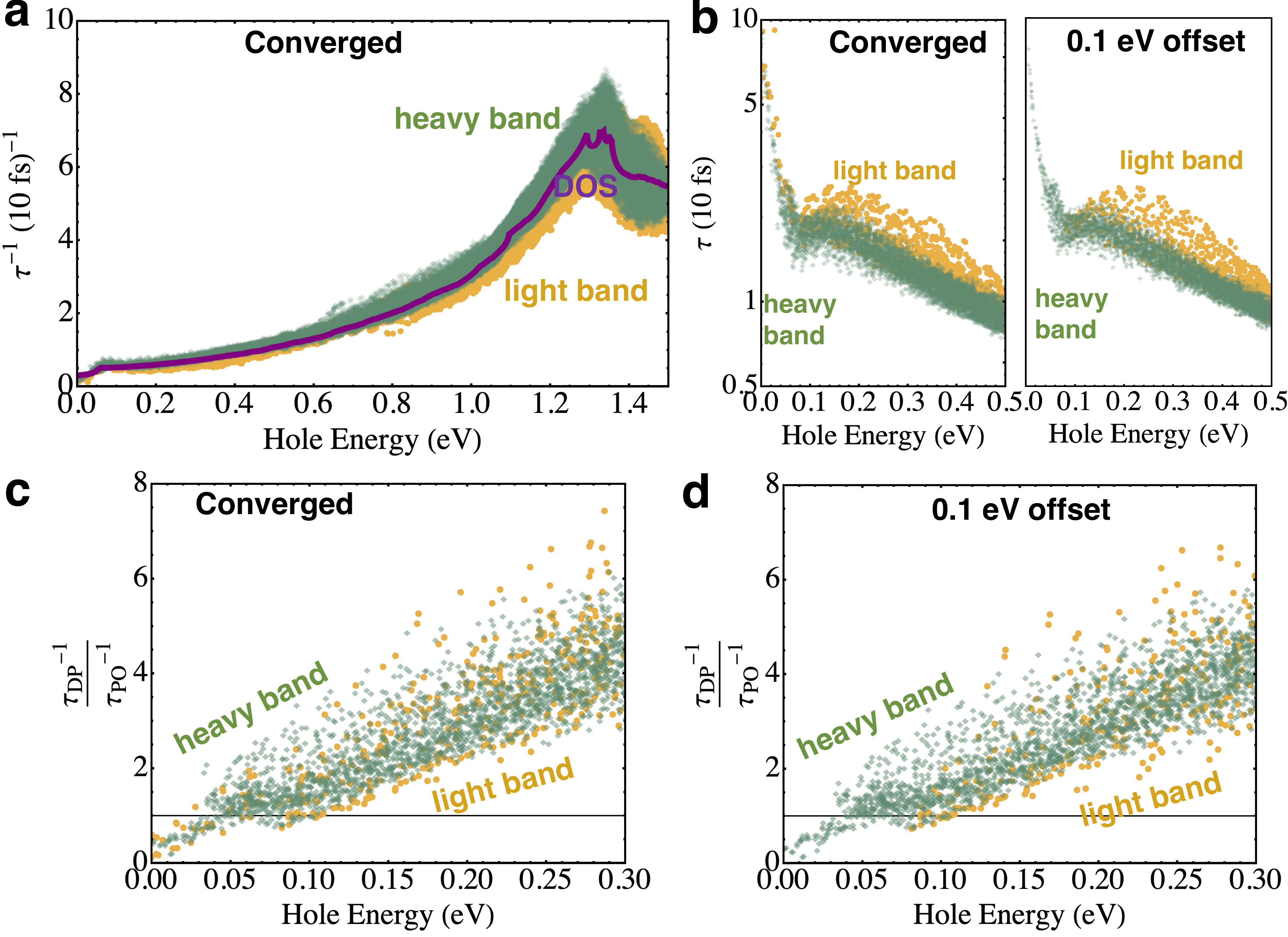}
\caption{ \textbf{Additional information on scattering in CaZnMgSb$_{2}$ at 600 K.} \textbf{a)} The overall scattering rates, where DOS is overlaid in purple. \textbf{b)} The overall hole lifetimes with band convergence and with 0.1 eV offset. \textbf{c)} The ratio of DPS to POS after convergence and \textbf{d)} before convergence with 0.1 eV offset.}
\label{fig:ephsupp}
\end{figure}

\newpage

\begin{figure}[hp]
\includegraphics[width=1 \linewidth]{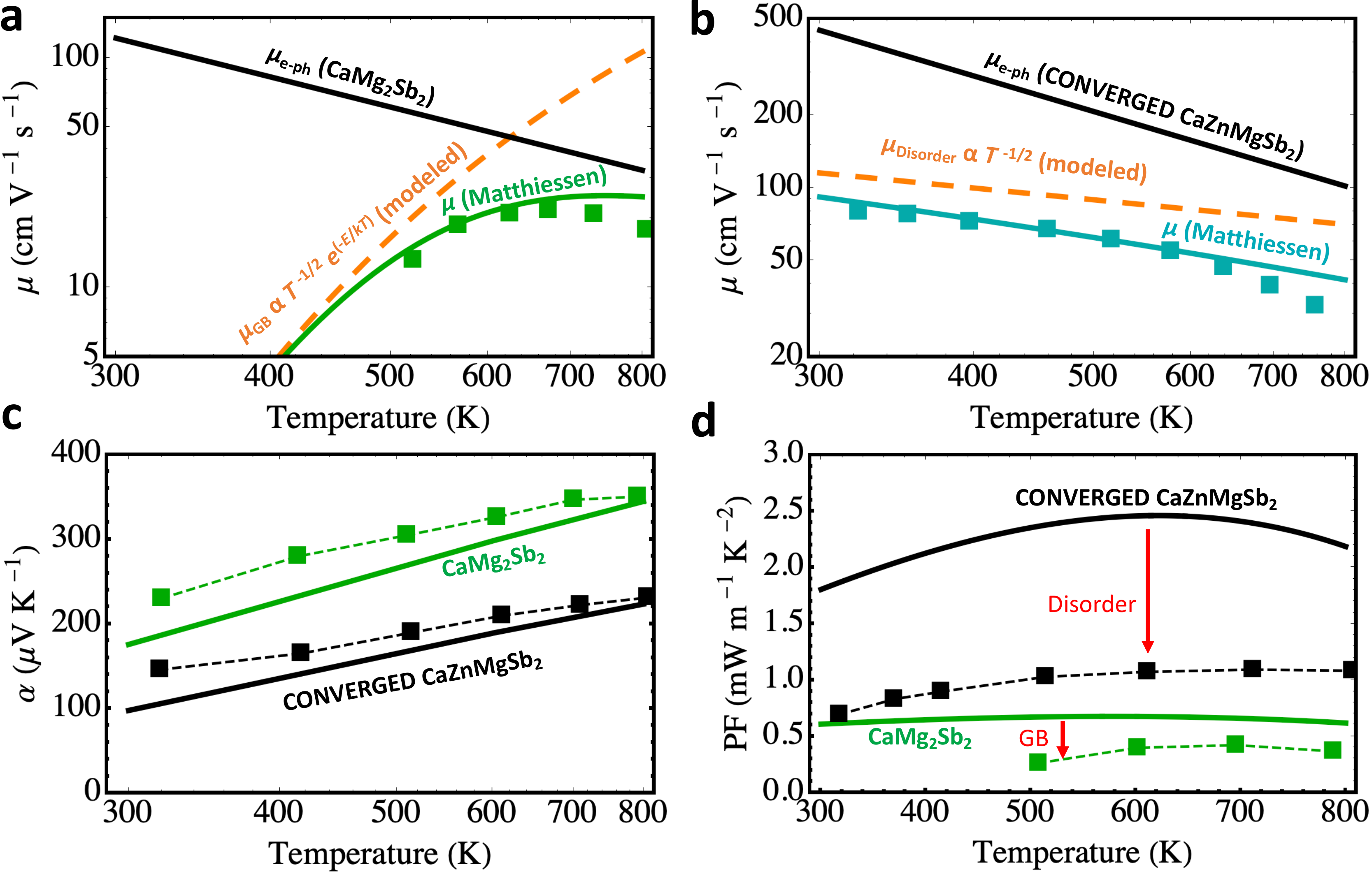}
\caption{ \textbf{Additional information on transport of CaZnMgSb$_{2}$ at 600 K.} \textbf{a--b)} Computed e-ph-limited mobilities, modeled grain-boundary-limited and disorder-limited mobilities, and the overall theoretical mobilities approximated by Matthiessen's rule, plotted against experimental data points at their respective experimental carrier concentration. \textbf{a)} CaMg$_{2}$Sb$_{2}$ at $n_{h}=10^{19}$ cm$^{-3}$. \textbf{b)} CaZnMgSb$_{2}$ at $n_{h}=3\times10^{19}$ cm$^{-3}$. \textbf{c)} Computed e-ph-limited Seebeck coefficients of  CaMg$_{2}$Sb$_{2}$ and band-converged CaZnMgSb$_{2}$ plotted against experimental data points (squares). \textbf{d)} Computed e-ph-limited power factors of CaMg$_{2}$Sb$_{2}$ and  band-converged CaZnMg$Sb_{2}$ plotted against experimental data points (squares).}
\label{fig:electronicsupp}
\end{figure}

\newpage

\begin{figure}[hp]
\includegraphics[width=0.6 \linewidth]{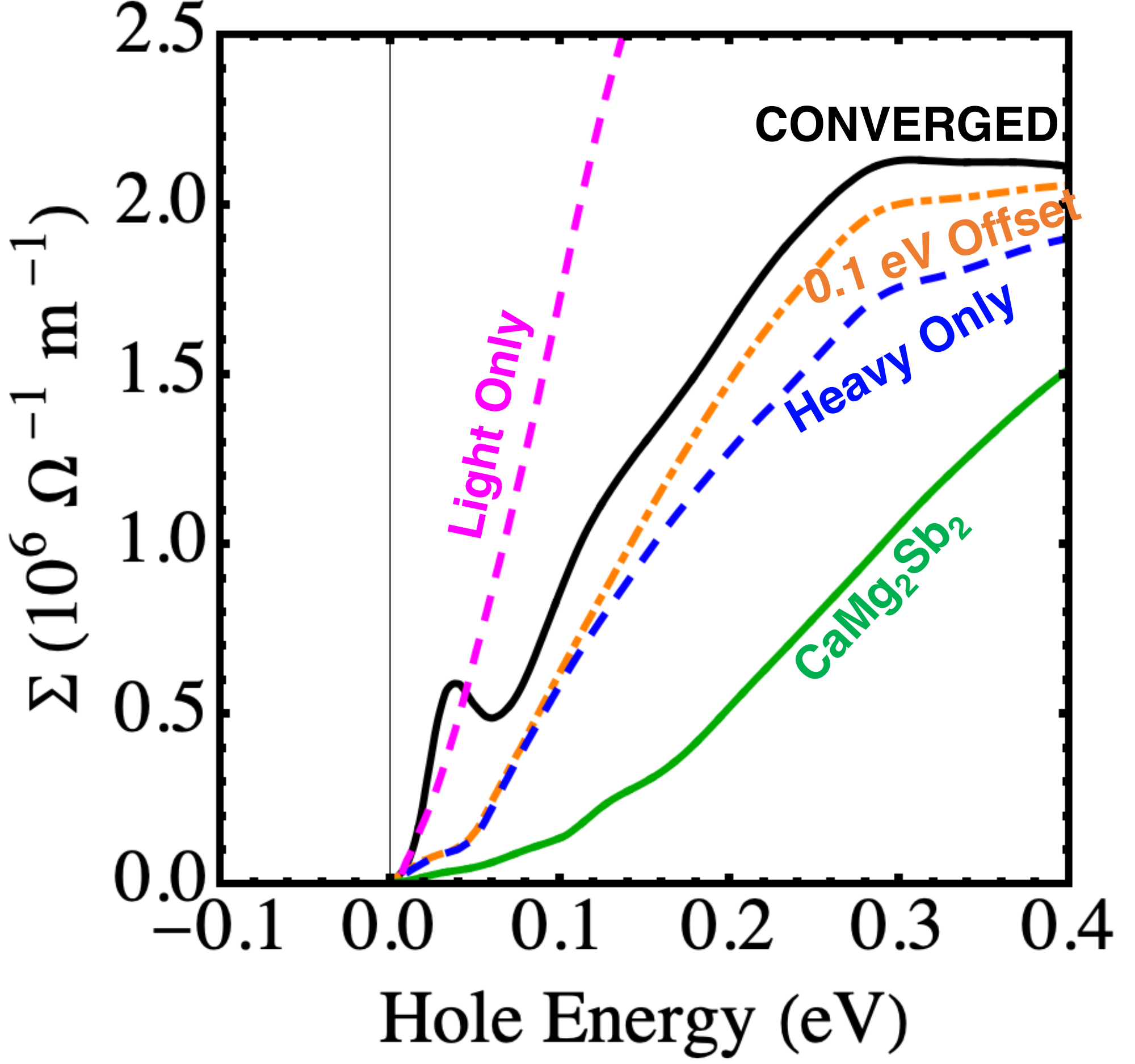}
\caption{ \textbf{Spectral conductivity $\Sigma(E)=v^{2}(E)\tau(E)D(E)$ of the band configurations at 600 K.}}
\label{fig:speccond}
\end{figure}

\newpage

\begin{figure}[hp]
\includegraphics[width=1 \linewidth]{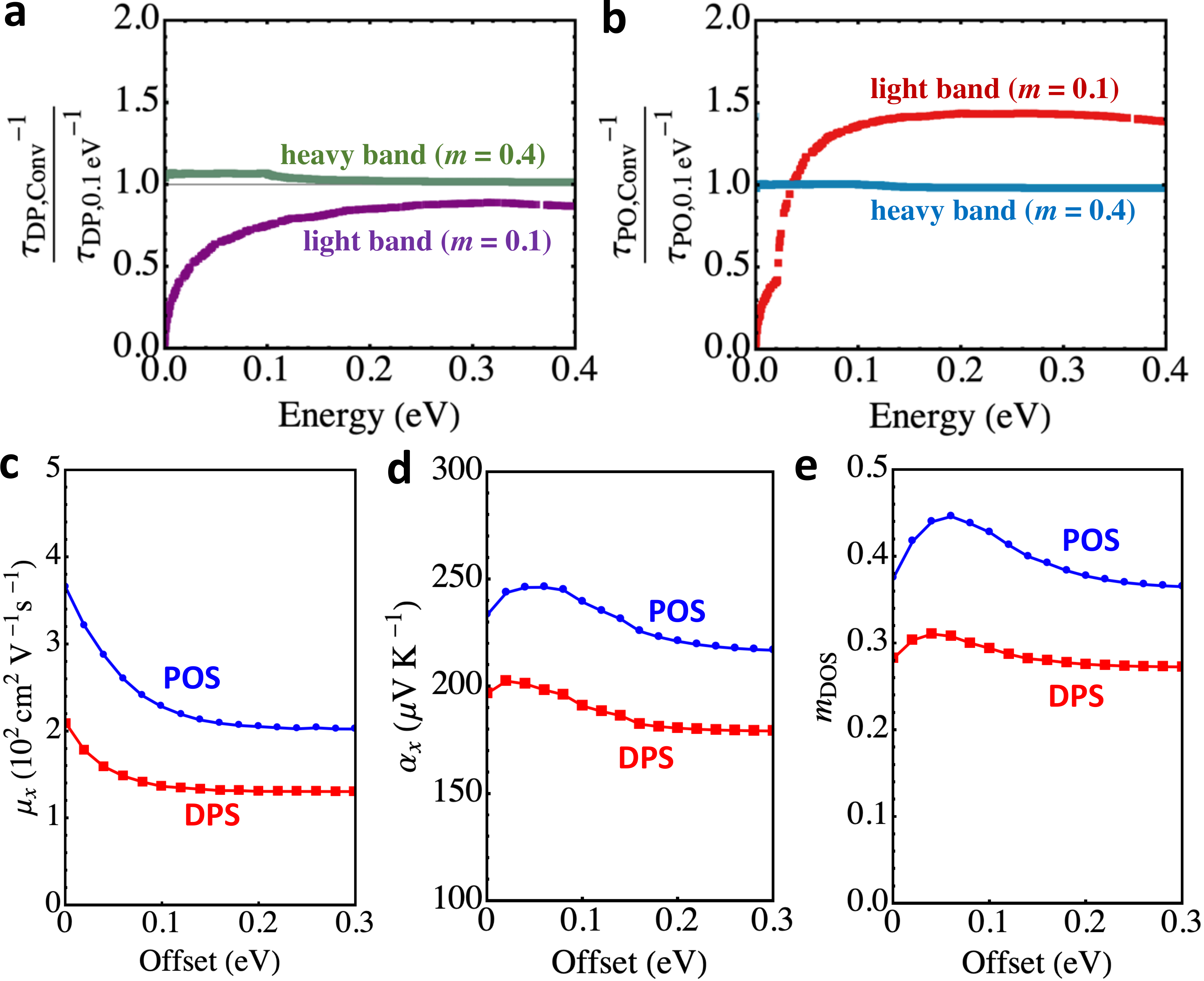}
\caption{ \textbf{Model scattering and transport properties at 600 K.} The model band structures and the scattering model are from Ref. \onlinecite{optimalbandstructure}. The band curvature effective masses of the upper and lower bands are 0.4 and 0.1 respectively, which are nearly equivalent to the corresponding values in CaZnMgSb$_{2}$ alloy. At each offset, the carrier concentration is optimized. \textbf{a)} The ratio of DPS and \textbf{b)} the ratio of POS before and after convergence by closing a 0.1 eV offset. \textbf{c)} Mobility increases with band convergence because the lighter band rises. \textbf{d)} The Seebeck coefficient maximizes short of full convergence. \textbf{e)} The DOS effective mass closely follows the Seebeck profile.}
\label{fig:modelscatteringzintl}
\end{figure}

\newpage

\begin{figure}[hp]
\includegraphics[width=1 \linewidth]{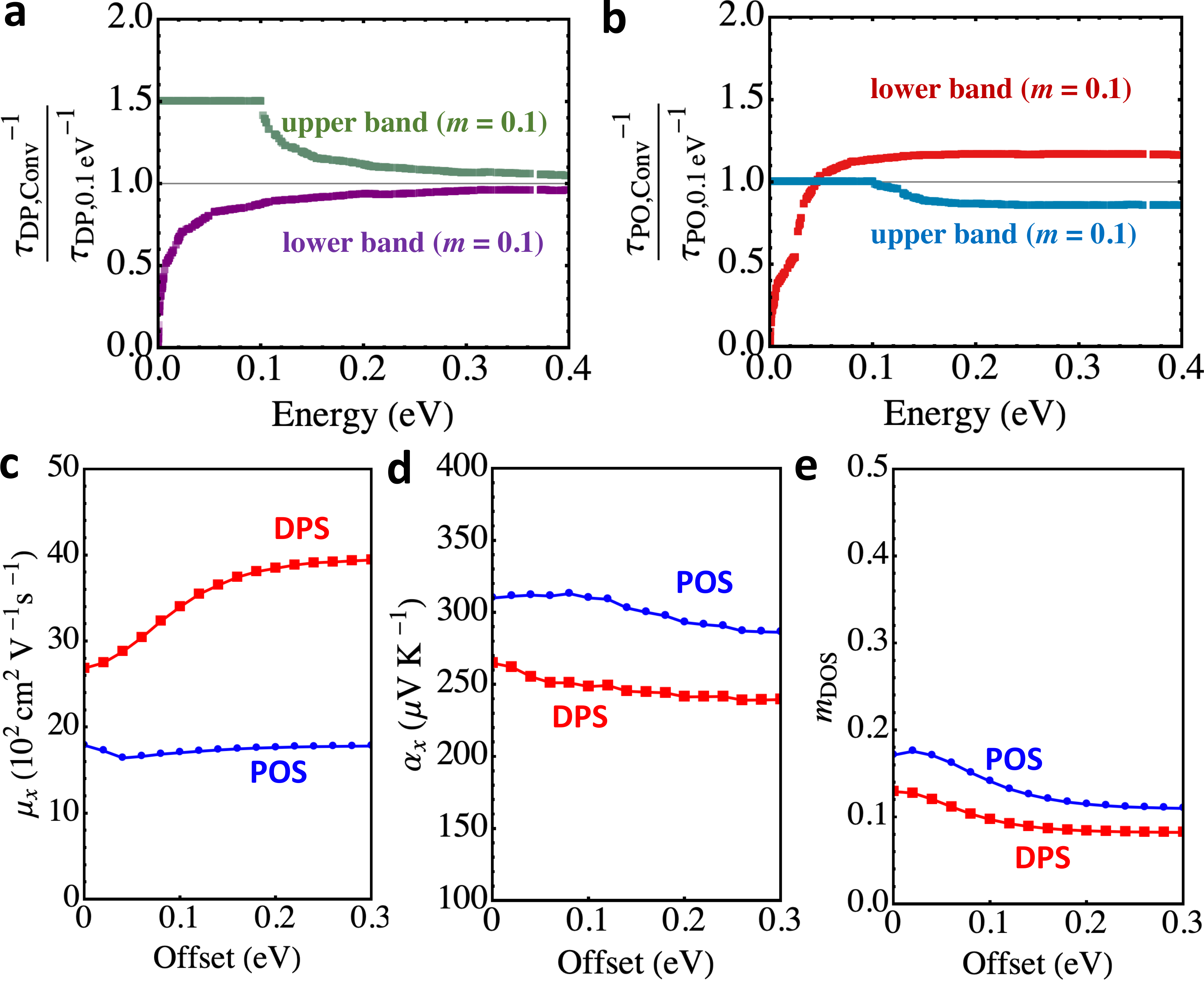}
\caption{ \textbf{Additional model scattering and transport properties at 600 K based on Ref. \onlinecite{optimalbandstructure}.} The two bands of identical curvature effective masses of 0.15. respectively. At each offset, the carrier concentration is optimized. \textbf{a)} The ratio of DPS and \textbf{b)} the ratio of POS before and after convergence by closing a 0.1 eV offset. \textbf{c)} Mobility decreases with band convergence in the DPS case because additional band means identical group velocity but heavier scattering due to interband transitions. \textbf{d)} The Seebeck coefficient essentially maximizes at full convergence. \textbf{e)} The DOS effective mass closely follows the Seebeck profile.}
\label{fig:modelscatteringsamem}
\end{figure}

\newpage

\begin{figure}[hp]
\includegraphics[width=0.7 \linewidth]{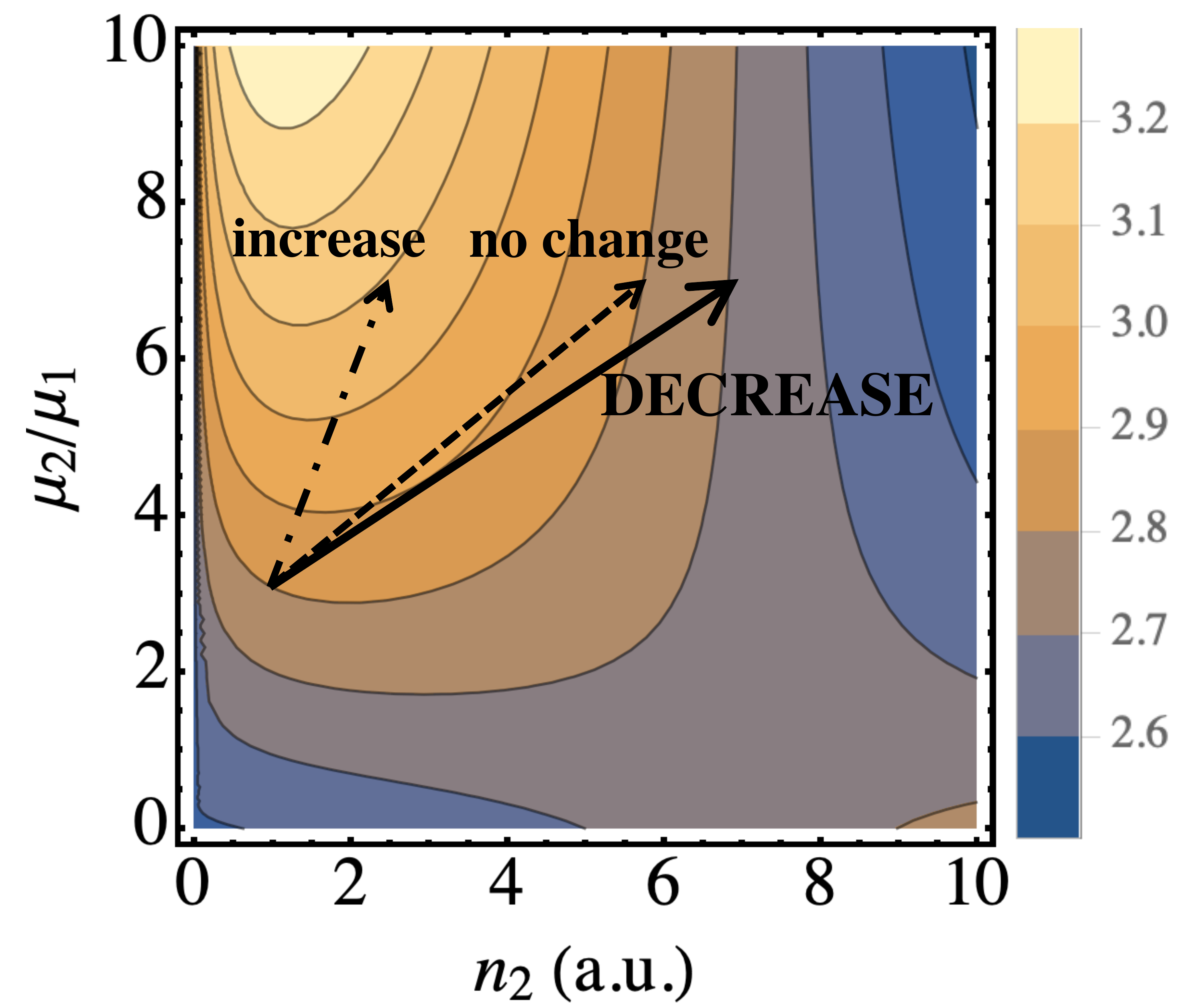}
\caption{ \textbf{Contour plot of the two-band Seebeck coefficient (Eq. \ref{eq:multiband2}) as a function of $n_{2}$ and mobility ratio $\mu_{2}/\mu_{1}$.} We choose parameters that closely represent the Zintl alloy case. Band curvature effective masses are chosen to be $m_{1}=0.4$ and $m_{2}=0.1$. Since lower band moves up by 0.1 eV, $n_{2}$ would increase approximately by a factor of 7 during convergence at 600 K. This path is traced by the solid arrow. Fixing $n= 77$ obeys the parabolic band relation $n_{1}/n_{2}=\left(m_{1}/m_{2}\right)^{3/2}=10$ at the point of convergence where $n_{1}=70$ and $n_{2}=7$. The dashed and dot-dashed arrows indicate other cases in which the Seebeck coefficient can increase or stay the same during band convergence.}
\label{fig:seebeckmodel}
\end{figure}

\newpage

\begin{figure}[hp]
\includegraphics[width=0.8 \linewidth]{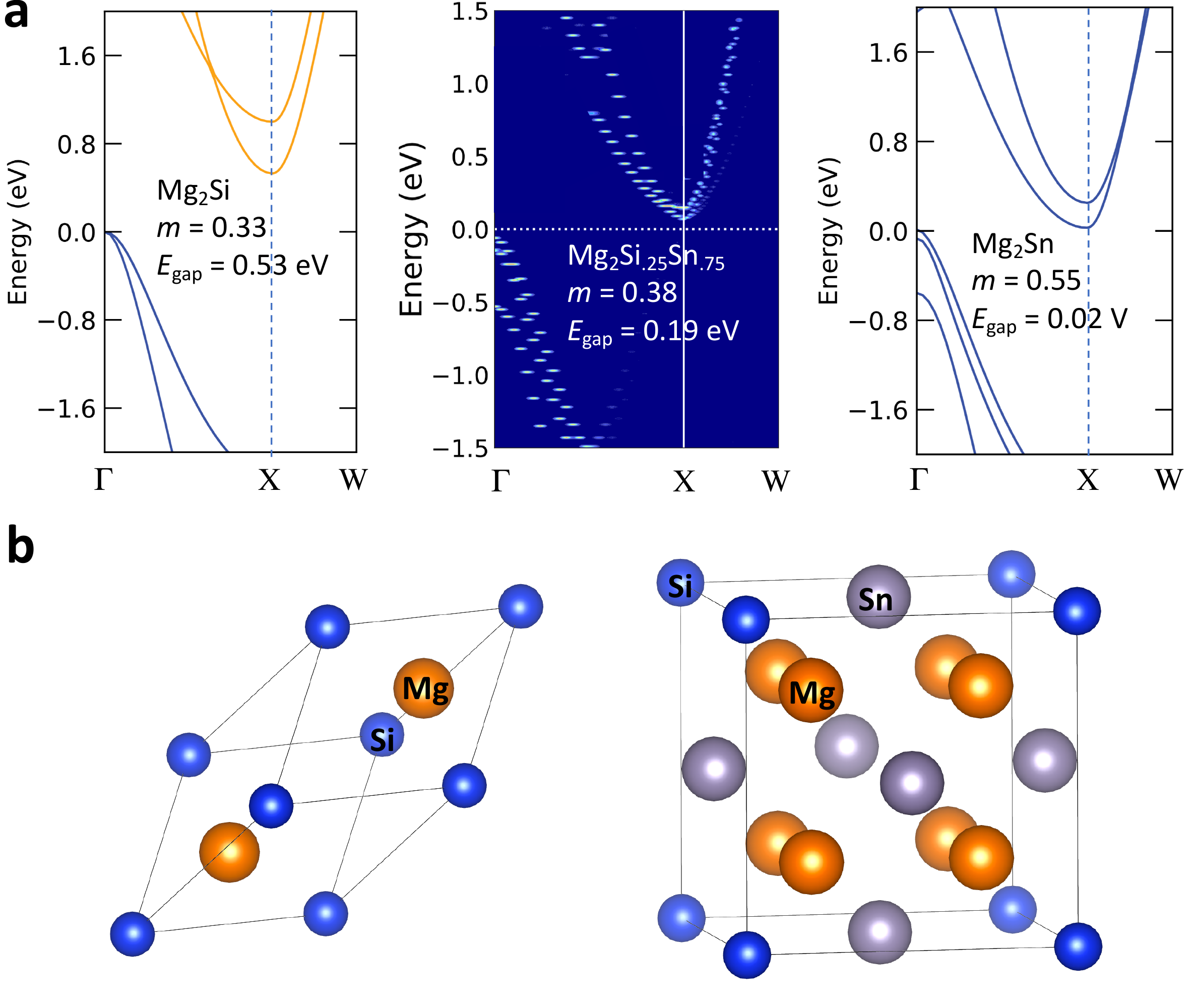}
\caption{ \textbf{The Mg$_{2}$Si$_{1-x}$Sn$_{x}$ system.} The alloys perform yield higher PF than either end compound and undergo band convergence at one \textbf{k}-point at around $x\sim0.7$. \textbf{a)} Band structures of the end compounds and Mg$_{2}$Si$_{0.25}$Sn$_{0.75}$. The alloy band structure is calculated using BandUP \cite{bandup1,bandup2}. Numbers in the figure are the harmonic mean of directional curvature effective masses and the band gap. \textbf{b)} The primitive crystal structures of Mg$_{2}$Si and Mg$_{2}$Si$_{0.25}$Sn$_{0.75}$. The latter is simple cubic whereas the former is face-centered cubic.}
\label{fig:mg2sisn}
\end{figure}

\newpage

\section{Supplementary Methods}
	
All formulae are provided in Hartree atomic units: $\hbar=m_{e}=e=4\pi\epsilon_{0}=1$

The matrix elements for scattering due to phonon deformation, the so-called ``deformation-potential" scattering, are computed by DFPT as follows:
\begin{equation}\label{eq:dpsmat}
g^{\text{DPS}}_{\nu^{\prime}\nu\lambda\mathbf{kq}}=\left\langle \psi_{\nu'\mathbf{k}+\mathbf{q}} \left|\sum_{ajl} \sqrt{\frac{1}{2M_{a}\omega_{\lambda\mathbf{q}}}} e^{i\mathbf{q}\cdot \mathbf{R}_{l}} \frac{\partial V}{\partial u^{al}_{\mathbf{q}j}}\hat{\mathbf{u}}^{al}_{\lambda\mathbf{q}j} \right|\psi_{\nu\mathbf{k}}\right\rangle
\end{equation}
where $\psi_{\nu}$ and $\psi_{\nu^{\prime}}$ are the initial and final electronic bands, $\partial V$ is the perturbation in the potential due to phonon deformation, {$\hat{\mathbf{u}}_{\lambda\mathbf{q}j}$} is phonon eigenvector for atom $a$, mode $\lambda$ and Cartesian direction $j$, and $u$ is atomic displacement, and $M_{a}$ is the mass of atom $a$. The polar-optical scattering matrix elements are computed as follows:
\begin{widetext}
\begin{equation}\label{eq:posmat}
g_{\nu'\nu\lambda\mathbf{kq}}^{\text{POS}}=i\sum_{a}\sqrt{\frac{1}{2M_{a}\omega_{\lambda\mathbf{q}}}}\sum_{\mathbf{G}\ne-\mathbf{q}} \frac{(\mathbf{q}+\mathbf{G})\cdot \mathbf{Z}^{*}\cdot\mathbf{\hat{u}}_{a\mathbf{\lambda\mathbf{q}}}}{(\mathbf{q}+\mathbf{G})\cdot\mathbf{\epsilon}^{\infty}\cdot(\mathbf{q}+\mathbf{G})} \left\langle \psi_{\nu'\mathbf{k}+\mathbf{q}} \right| e^{i(\mathbf{q}+\mathbf{G})\cdot(\mathbf{r-\textit{\textbf{l}}})} \left| \psi_{\nu\mathbf{k}}\right\rangle,
\end{equation}
\end{widetext}
where $\mathbf{Z}^{*}$ is the Born effective charge tensor, $\mathbf{\epsilon}^{\infty}$ is the high-frequency dielectric permittivity tensor, \textbf{G} is the reciprocal lattice vector, and \textbf{r} is atomic position vector. The total matrix elements are sums of the two: $g=g^{\text{DPS}}+g^{\text{POS}}$. Fig. \ref{fig:decay} validates accurate Wanniner interpolation and proper localization of the interpolated quantities (band structure, phonon dynamical matrix, and the scattering matrix).

As the two original Mg sites are symmetry-equivalent, it matters not which one is occupied by Zn for CaZnMgSb$_{2}$. We omit calculation of the CaZn$_{2}$Sb$_{2}$ because it requires an advanced functional to capture its semiconducting character and therefore is not amenable to EPW. See Fig. \ref{fig:cazn2sb2} for its metallic PBE+SOC band structure.

\section{Supplementary Discussions}

\subsection{On the Seebeck Coefficient and Density-of-States Effective Mass}

It may seem unintuitive at first that the DOS effective mass ($m_{\text{D}}$) decreases with band convergence, as convergence of bands increases the total DOS. We emphasize here that $m_{\text{D}}$ is not uniquely determined by DOS or band profile, and is certainly not guaranteed to monotonically increase with the total DOS. Even if two bands belonging to two different systems had identical DOS and band curvature, their $m_{\text{D}}$ could still be different if the dominant scattering mechanism were different between the two systems. Because $m_{\text{D}}=\left( \frac{\mu_{w}}{\mu} \right)^{2/3}$, and weighted mobility ($\mu_{w}$) is defined through the Seebeck coefficient ($\alpha$) as per Eqs. 1--2  in the main text, it is the relative behaviors of $\alpha$ and $\mu$ that ultimately determine $m_{\text{D}}$. If scattering changes in behavior such that $\mu$ decreases and $\alpha$ increases, then $m_{\text{D}}$ increases. If $\mu$ increases but $\alpha$ decreases, then $m_{\text{D}}$ decreases.

It is a well known consequence that the Seebeck coefficient of a given band changes with the scattering mechanism. In the single parabolic band model \cite{nolassharpgoldsmid}, 
\begin{equation}\label{eq:spbseebeck}
\alpha=k_{\text{B}}\left( \frac{E_{\text{F}}}{k_{\text{B}}T}-r-\frac{5}{2}\right)
\end{equation}
where $r$ is the power of energy to which scattering rates scale. For instance, if $\tau^{-1}\propto E^{0.5}$ under one scattering mechanism whereas $\tau^{-1}\propto E^{-0.5}$ under another, $\alpha$ is higher under the latter scattering regime. Even if the overall lumped scattering rate were equal between the two processes such that they generate the same $\mu$, because the energy-dependence of their rates remain different, $\alpha$ will be higher under the latter. Therefore, $\mu_{w}$ and $m_{\text{D}}$ are both higher under the latter.

Now, let us first consider the typical manner in which band-convergence is conceived, where interband scattering is minimal or inexistent. In this case, additional pockets generally increase $\alpha$ at a given carrier concentration, and $\mu$ is nominally not altered due to band convergence. It follows then that increase in total DOS indeed leads to increase in $m_{\text{D}}$, as expected. 

When opportunity for interband scattering is present, then scattering rates can no longer generally be modeled with a simple power of energy. It nevertheless remains true that certain scattering behaviors would be more beneficial for $\alpha$ than others. Interband scattering can change in behavior in a way $\alpha$ could still increase but also decrease, which would be mirrored by $\mu_{w}$ and $m_{\text{D}}$. In our Zintl case, as the light band arises to converge the heavy band at the valence band maximum, it increases $\mu$ (higher overall group velocity) but incurs changes in DPS and particularly POS that reduces $\alpha$. Combination of higher $\mu$ but lower $\alpha$ and hence stagnant $\mu_{w}$ leads to decreased $m_{\text{D}}$, in spite of increased total DOS. In short, whether the total DOS increases or decreases is irrelevant for $m_{\text{D}}$, at least not directly, and affects $m_{\text{D}}$ only through what it means for interband scattering.

To graphically portray the impact interband scattering has on $\alpha$, we provide in Fig. \ref{fig:speccond} comparative spectral conductivitity of the Zintl band configurations. The profile of $\Sigma(E)=v^{2}(E)\tau(E)D(E)$ is what determines $\alpha$, as dictated by Boltzmann transport formalism
\begin{equation}\label{eq:speccond}
\alpha=\frac{1}{T}\frac{\int \Sigma(E)(E_{\text{F}}-E)\left(-\frac{\partial f}{\partial E}\right)dE}{ \int \Sigma(E)\left(-\frac{\partial f}{\partial E}\right)dE},
\end{equation}
where one may identify the denominator with Ohmic charge conductivity ($\sigma$). The only difference between the numerator and the denominator is the term $E_{\text{F}}-E$ in the former, which stipulates that for $\alpha$ to be high, $\Sigma(E)$ should be relatively higher further away from the Fermi level (i.e. higher energies) to increase the numerator, and relatively lower near the Fermi level (i.e. lower energies) to decrease the denominator. Of note, this is the underlying reason why a smaller value of $r$ is beneficial in Eq. \ref{eq:spbseebeck} for single parabolic band, i.e., it is more beneficial for lifetimes to be higher (scattering to be weaker) at higher energies in relative terms. Also for these reasons has the rule-of-thumb that steeply increasing $\Sigma(E)$ benefits $\alpha$ has come to be. This, however, presumes that scattering rates (hence lifetiems) are simply behaved such that $\Sigma(E)$ increases monotonically with $E$. This presumption is not generally valid in the presence of interband scattering, as evidenced by Fig. \ref{fig:speccond} 

We in fact find that $\Sigma(E)$ of the band-converged configuration does not behave trivially. It starts out with an even steeper onset than that of the lower-band-only case owing to coexistence of the two bands supplying higher carrier concentration. However, it reverses course and dips around 0.07 eV corresponding to the portion of the light band that increases in overall scattering due to convergence, which is slightly past the emission onset for POS. Emission onset is the  energy below which inelastic phonon emission is very weak if not outright forbidden due to absence of final states (e.g. electron at 30 meV cannot emit a phonon with a frequency of 50 meV because -20 meV is in the band gap). Thus, emission onset is usually close in value to the highest optical frequency. Thereafter, POS is fully activated, and $\Sigma(E)$ again increases with $E$ but this time with a more gradual slope - similar to those of the 0.1-eV-offset and heavy-band-only cases, indicating that the heavy band dominates transport from there on. From the perspective of Eq. \ref{eq:speccond}, this $\Sigma(E)$ profile of the band-converged case is not the most favorable for $\alpha$. The protrusion at $E<0.07$ eV increases the denominator of Eq. \ref{eq:speccond}, improving $\sigma$ and hence $\mu$ over the other band configurations. However, the dip that follows reduces the numerator of Eq. \ref{eq:speccond} relative to the denominator, reducing $\alpha$.

We further support our observation by means of complementary modeling studies. Fig. \ref{fig:modelscatteringzintl} is based on Ref. \onlinecite{optimalbandstructure}, whereby realistic model band structures and modifications of analytic scattering models have been used. We use two bands with $m=0.4$ and $m=0.1$, whether the former is fixed atop and the latter shifts for convergence, for the semblance with the CaZnMgSb$_{2}$ bands in question. In Figs. \ref{fig:modelscatteringzintl}a--b, the scattering rates are computed at 600 K for bands. Under DPS, the lower band only decreases and the upper band only increases in scattering due to convergence. Under POS, the lower band increases in scattering due to convergence past the emission onset. The results establish that the scattering behavior of DPS and POS during convergence computed for CaZnMgSb$_{2}$, in Fig. 3 of the main text, are generalizable. Figs. \ref{fig:modelscatteringzintl}c--e qualitatively reproduce that $\alpha$ and $m_{\text{D}}$ do not peak at full convergence but rather around 0.1 eV, even though $\mu$ continues to improve up to full convergence. Because $\alpha$ is different under the two scattering regimes, $m_{\text{D}}$ is also different under the two regimes (through $\mu_{w}$). That is, $m_{\text{D}}$ is not uniquely determined by the band structure or DOS alone.

If as in Fig. \ref{fig:modelscatteringsamem} two identical bands are used instead, each with $m=0.15$ and $m=0.15$, then the behaviors are somewhat different. In this case, even with interband scattering. $\alpha$ continues to increase with convergence until full convergence as well as $m_{\text{D}}$. These behaviors are consistent with the way in which band convergence is typically conceived. The underlying reason is that when the two bands share the same shape, then interband scattering cleanly increases by a constant factor upon full convergence, as opposed to when two bands have disparate curvature profiles. Nevertheless, $\mu$ decreases because the bands' identical shapes do not change group velocity while inviting more scattering. Of course, $\alpha$ and $m_{\text{D}}$ still depend on the scattering regime.

In short summary, when two bands are identical, decrease in $\mu$ is the main limiting agent of the benefit of band convergence since $\alpha$ is expected to improve. On the other hand when a lighter band converges upon a heavier band, then decrease in $\alpha$ is the limiting agent while $\mu$ is expected to improve. This is the case for the Zintl alloy. If a heavier band converges upon a lighter band, the band convergence is likely to be harmful since the overall group velocity drops in addition to increased scattering.

Even more qualitative analysis can also establish the possible reduction of $\alpha$ due to band convergence. Given two bands, the Seebeck coefficient can be expressed as
\begin{equation}\label{eq:multiband1}
\alpha=\frac{\sigma_{1}\alpha_{1}+\sigma_{2}\alpha_{2}}{\sigma_{1}+\sigma_{2}}=\frac{n_{1}\mu_{1}\alpha_{1}+n_{2}\mu_{2}\alpha_{2}}{n_{1}\mu_{1}+n_{2}\mu_{2}},
\end{equation}
where $n$ denotes carrier concentration and subscript 1 and 2 respectively indicate the upper band (fixed) and the lower band (converging). Using the Seebeck coefficient expression for a degenerately doped single parabolic band,
\begin{equation}\label{eq:spbalpha}
\alpha=\frac{2k_{\text{B}}^{2}T}{3}m\left(\frac{\pi}{3n}\right)^{\frac{2}{3}},
\end{equation}
we can recast Eq. \ref{eq:multiband1} as follows:
\begin{equation}\label{eq:multiband2}
\begin{aligned}
\alpha&=\frac{2k_{\text{B}}^{2}T}{3}\left(\frac{\pi}{3}\right)^{\frac{2}{3}}\frac{m_{1}n^{\frac{1}{3}}_{1}\mu_{1}+m_{2}n^{\frac{1}{3}}_{2}\mu_{2}}{n_{1}\mu_{1}+n_{2}\mu_{2}} \\
&=\frac{2k_{\text{B}}^{2}T}{3}\left(\frac{\pi}{3}\right)^{\frac{2}{3}}\frac{m_{1}(n-n_{2})^{\frac{1}{3}}+m_{2}n^{\frac{1}{3}}_{2}\left(\frac{\mu_{2}}{\mu_{1}}\right)}{(n-n_{2})+n_{2}\left(\frac{\mu_{2}}{\mu_{1}}\right)}.
\end{aligned}
\end{equation}
Then for given $m_{1}$, $m_{2}$, and fixed total $n$ ($n$=$n_{1}+n_{2}$), the Seebeck coefficient can be modeled after Eq. \ref{eq:multiband2} by using $n_{2}$ and $\mu_{2}/\mu_{1}$ as two tunable parameters. The result is illustrated in Fig. \ref{fig:seebeckmodel}. As the lower band rises to converge with the upper band at the band edge, $n_{2}$ increases. Also, because the lower band decreases in scattering while the upper band increases in it, $\mu_{2}/\mu_{1}$ increase. In the Zintl case, convergence by 0.1 eV increases $n_{2}$ by about sevenfold. The arrows in Fig. \ref{fig:seebeckmodel} show that, for given increase in $n_{2}$ (by sevenfold) and given increase in $\mu_{2}/\mu_{1}$ (by threefold), it is possible for the overall $\alpha$ to decrease in spite of increased $\mu_{2}/\mu_{1}$ depending on what the values of $\mu_{2}/\mu_{1}$ are before and after convergence. If the $\mu_{2}/\mu_{1}=1$ after convergence, meaning that the two bands are identical, then $\alpha$ increases due to their convergence (arrow 1). For some value that is $\mu_{2}/\mu_{1}>1$, where the lower band is somewhat lighter than the upper band, band convergence does not change $\alpha$ (arrow 2). But if $\mu_{2}/\mu_{1}>>1$, meaning that the lower band is far lighter than the upper band, then band convergence can reduce $\alpha$ (arrow 3).

\subsection{More on Carrier Scattering and Transport of the Zintl alloys}

Fig. \ref{fig:ephsupp} shows that while scattering overall mostly follows the DPS trend of $\tau^{-1}\propto$ DOS, the states within 0.1 eV of the band edge are affected by POS. This is a wide enough width to affect the Seebeck coefficient behavior.

As alloy scattering, grain-boundary scattering, and ionized-impurity scattering were not accounted for in computation, our general overestimation of experimental $\mu$, $\mu_{w}$, $\sigma$, and the PF comes with no surprise. Their absence must be causing some discrepancies in $\alpha$ as well. Comparisons with experimental data are provided in Fig. \ref{fig:electronicsupp}. 

Mg$_{3}$Sb$_{2}$-like Zintl compounds are known to experience significant grain-boundary (GB) scattering that manifests in the form of temperature-activated, hopping-like electronic transport behavior at lower temperatures \cite{mg3sb2grainboundary}. The CaMg$_{2}$Sb$_{2}$ samples were indeed highly polycrystalline, which explains the skewed temperature-dependence of mobility in Figs. \ref{fig:electronicsupp}a. Reduction of intrinsic $\mu$ by half or more due to GB scattering is routinely reported for alloy thermoelectrics \cite{chalcogenidegrainboundary,srtio3grainboundary,mg3sb2grainboundary,mg2sigrainboundary}. It is well known that electronic transport through GB is a temperature-activated process trending generally as $\mu_{\text{GB}}\propto T^{-0.5}\text{exp}\left(\frac{-\Phi_{\text{GB}}}{k_{\text{B}}T}\right)$ where $\Phi_{\text{GB}}$ is the potential barrier at the grain boundary \cite{chalcogenidegrainboundary,snsegrainboundary,mg2sigrainboundary}. When we account for this model with parameter adjustment, we see that the experimentally measured mobility behavior of CaMg$_{2}$Sb$_{2}$ is recovered reasonably well. The remaining discrepancy at high temperatures indicates either imperfect model specification for GB scattering or our neglect of ionized-impurity scattering that would arise in experiments due to the Na dopants. 

The alloy for which the full band-convergence was achieved, CaZn$_{1.14}$Mg$_{0.86}$Sb$_{2}$, is not only stoichiometrically non-integral but also likely possesses high levels of site disorder. This would introduce disorder scattering that are absent in the perfectly ordered CaZnMgSb$_{2}$ on which calculations are run. Disorder scattering for electrons is generally modeled as $\mu_{\text{disorder}}\propto T^{-0.5}$ \cite{hgtedisorderscattering}. Using this model with suitable parameter adjustment and then applying Matthiessen's rule, we recover the experimental mobility reasonably well as shown in Fig. \ref{fig:electronicsupp}b. At 600 K, disorder roughly halves intrinsic e-ph mobility. The remaining discrepancy at high temperatures again indicates either imperfect model specification for disorder scattering, our neglect of other scattering processes such as ionized-impurity scattering from dopants.

Reduction of intrinsic $\mu$ by half or more due to disorder scattering is also routinely reported. In Mg$_{2}$Si$_{1-x}$Sn$_{x}$ alloys, experimentally fitted effective deformation potential (which lumps the effects of DPS of all kinds: acousiic, optical, intraband and interband \cite{wangsnyderbook}) of 13 eV and alloy scattering potential of 0.7 eV corresponded to roughly equal degree of scattering between the two, with one roughly halving the mobility limited by the other \cite{mg2sibandconvergence1}. In comparison in the present Zintl alloys, the experimentally inferred scattering potentials were respectively 10 eV and 0.5 eV \cite{zintlvalencecrossing}, which indicate similar relative strengths of the two mechanisms.

\subsection{On M\lowercase{g}$_{2}$S\lowercase{i}$_{1-x}$S\lowercase{n}$_{x}$}

As mentioned in the main text, Mg$_{2}$Si$_{1-x}$Sn$_{x}$ is a system whose $n$-type PF clearly benefits from alloying, peaking around $x\sim0.7$ \cite{highlyeffectivemg2sisn,mg2sibandconvergence1,mg2sibandconvergence2}. It is presumed that the convergence of its two lowermost conduction bands is responsible for this enhancement, though it occurs at one point (the $X$-point). We confirm that this convergence occurs by calculating of alloy band structure at $x=0.75$ using HSE06 \cite{hse1,hse2} hybrid functional and BandUP \cite{bandup1,bandup2}, shown in Fig. \ref{fig:mg2sisn}. However, we cannot positively confirm here whether band convergence indeed is the (primary) reason because the alloys are metallic at the plain DFT level even with $x\sim0.5$, and hence not amenable to an EPW study. That said, we find a few other reasons that could be primarily responsible for the peak PF around $x\sim0.7$.

First, the band gap substantially decreases towards the Sn-end ($x=1$). As computed with the HSE06 hybrid functional \cite{hse1,hse2} with SOC, the band gap is 0.53 eV for Mg$_{2}$Si, 0.19 eV for Mg$_{2}$Si$_{0.25}$Sn$_{0.25}$ (which is already fairly small) where the bands are close to converging, and 0.02 eV for Mg$_{2}$Sn (essentially semimetallic). Therefore, increasing bipolar effect is likely a big reason that the PF falls toward the Sn-end as $x$ increases beyond some optimal value.

Second, the reason that the PF increases with $x$ from the Si-end ($x=0$) toward $x\sim0.7$ may at least partly have to do with the fact that, in Mg$_{2}$Si, the converging higher-energy band is more favorable for thermoelectrics than the lower-energy band. The latter's profile is approximately $m_{\parallel}=0.19$ and $m_{\perp}=0.58$ for $\frac{m_{\perp}}{m_{\parallel}}=3.1$, while the former's profile is $m_{\parallel}=0.13$ and $m_{\perp}=0.97$ for $\frac{m_{\perp}}{m_{\parallel}}=7.5$. The former has a smaller curvature mass in the light direction as well as higher degree of anisotropy, both of which benefit thermoelectricity. As $x$ continues to increase, this band continues to shift downward, which would likely improve the PF perhaps even unto the Sn-end if not for bipolar effect.

Third, as mentioned in the previous section, the experimentally fitted effective deformation potential for Mg$_{2}$Si$_{1-x}$Sn$_{x}$ was greater than that of CaZn$_{2-x}$Mg$_{x}$Sb$_{2}$, at 13 eV against 10 eV. This means that, unless intraband deformation potential is significantly stronger in Mg$_{2}$Si$_{1-x}$Sn$_{x}$ than in the Zintl system, it is unlikely that interband scattering is relatively weaker in Mg$_{2}$Si$_{1-x}$Sn$_{x}$, and therefore possibly just as ineffective for improving thermoelectric performance.

Of course, it remains entirely possible that interband scattering is simply weaker than seen in the Zintl system, only it is challenging to positively confirm. Moreover, the two bands are not as disparate in shapes as the two bands in the Zintl system, which bodes better for thermoelectric performance once they converge in spite of interband scattering. In these cases, band convergence would play a major part in performance improvement in the Mg$_{2}$Si$_{1-x}$Sn$_{x}$ alloys. 

In all, there are likely many different kinds of changes made to the electronic structure of the system when the alloy is created experimentally, clouding whether energy convergence of the bands is the primary let alone the sole reason that the performance peaks at some intermediate $x$, until it can be directly tested.

\section{Supplementary References}

\bibliography{references}